\documentclass[11pt,a4paper]{article}

\usepackage{fontspec}
\usepackage{amsmath}
\usepackage{amssymb}
\usepackage{graphicx}
\usepackage{booktabs}
\usepackage{longtable}
\usepackage{array}
\usepackage{caption}
\usepackage{calc}
\usepackage[margin=2.4cm]{geometry}
\usepackage[hidelinks]{hyperref}
\usepackage{newunicodechar}
\usepackage{etoolbox}
\usepackage{authblk}

\providecommand{\tightlist}{\setlength{\itemsep}{0pt}\setlength{\parskip}{0pt}}

\providecommand{\real}[1]{#1}

\AtBeginEnvironment{longtable}{\footnotesize\linespread{1.0}\selectfont
  \setlength{\tabcolsep}{4pt}}

\newunicodechar{α}{\ensuremath{\alpha}}
\newunicodechar{β}{\ensuremath{\beta}}
\newunicodechar{γ}{\ensuremath{\gamma}}
\newunicodechar{ρ}{\ensuremath{\rho}}
\newunicodechar{∈}{\ensuremath{\in}}
\newunicodechar{±}{\ensuremath{\pm}}
\newunicodechar{²}{\textsuperscript{2}}
\newunicodechar{–}{\textendash}
\newunicodechar{−}{\ensuremath{-}}

\setkeys{Gin}{width=0.86\linewidth,keepaspectratio}
\title{\bfseries From Evaluated Models to Evaluation Aids:\\[2pt]
A Multi-Evidence Study of LLM-Based Difficulty Calibration\\ for Programming Examinations}

\author[1,4,5]{Hongfei Yan}
\author[2]{Jiangkai Xiong}
\author[2]{Yiqing Li}
\author[3]{Chong Chen}
\affil[1]{School of Computer Science, Peking University, Beijing 100871, China}
\affil[2]{Yuanpei College, Peking University, Beijing 100871, China}
\affil[3]{School of Government, Beijing Normal University, Beijing 100875, China}
\affil[4]{National Key Laboratory for Multimedia Information Processing, Beijing 100871, China}
\affil[5]{Beijing Key Laboratory of AI Systems, Beijing 100871, China}
\date{}

\begin{document}
\maketitle

\begin{abstract}
\noindent Difficulty differences across parallel-class programming examinations directly affect the fairness of course assessment. This study repositions large language models from evaluation targets in code-generation benchmarks to auxiliary evidence sources for interpreting exam difficulty, and builds a multi-evidence framework combining AI evidence, aggregated student performance, item exposure, online-judge process data, and teacher interpretation. In a first-stage experiment, ten models solved an eight-problem final examination synchronously with 120 students: AI pass rate correlated positively with student pass rate (Spearman rho = 0.866, exact two-sided permutation p = 0.0119), and the solving-based composite difficulty index correlated negatively with student pass rate (rho = -0.905, exact p = 0.0046). Building on this, a single structured reviewer was run through provenance-preserving API calls (a third-party OpenAI-compatible endpoint whose requested and returned model label, gpt-5.6-sol, cannot authenticate an official OpenAI upstream model; endpoint, temperature, prompt/schema hashes, and raw responses are archived). In a cross-sectional sample of 79 problems from 11 parallel-class final examinations, AI overall difficulty correlated with student problem-level pass rate at Spearman rho = -0.871 and with non-attempt rate at rho = 0.800; in a 26-problem longitudinal Data Structures and Algorithms B sample from the same instructor, the corresponding correlations were -0.829 and 0.883. A 106-problem introductory-course (CS101) sample marks the boundary of the scale: the problem-level correlation weakened to rho = -0.552, and the exam-level correlation across 16 examinations was near zero, with cohort composition rather than problem difficulty dominating exam-level outcomes. Exposure-discount sensitivity analysis (0-0.40) and duplicate-problem perturbation tests did not change the direction of these correlations. The findings indicate that solving-based and review-based AI evidence can serve as external references for problem validation, parallel-class fairness discussion, and longitudinal quality tracking at the problem and exam-structure level, while the third-party endpoint's model-identity boundary, the single-reviewer design, and review-output instability define explicit limits of use: AI difficulty scales must not be used for individual student evaluation or automatic grade adjustment.

\vspace{0.8em}
\end{abstract}

\hypertarget{introduction}{%
\section{Introduction}\label{introduction}}

Programming courses are a core foundational course in the training of computer-related talent at universities. As course enrollment has expanded, a single course is often offered by multiple teachers in parallel classes, which organize training, item setting, and examinations separately under a unified syllabus. Online Judge (OJ) platforms have gradually evolved from early experimental systems into mature tools for large-scale automated scoring, supporting objective evaluation and process recording in programming education {[}1{]}{[}2{]}{[}3{]}{[}4{]}; however, they have not thereby resolved the problem of comparability of exam difficulty across parallel classes. Different teachers make differing judgments regarding knowledge-point coverage, algorithmic depth, implementation complexity, boundary data, problem statements, and time limits, which leaves student scores without a stable reference point in cross-class comparison, course-quality analysis, and instructional improvement.

Policy documents such as the ``Overall Plan for Deepening Educational Evaluation Reform in the New Era'' emphasize improving outcome evaluation, strengthening process evaluation, exploring value-added evaluation, and refining comprehensive evaluation, and they propose using modern information technologies such as artificial intelligence and big data to innovate evaluation tools {[}5{]}{[}6{]}. Programming examinations have inherently digital conditions for evaluation: problems can be judged by machines, submission processes can be recorded, and AI models can also generate, submit, and revise programs under the same problem statement and judging environment. This makes ``having AI participate in exam difficulty calibration'' an operable research question.

However, existing research on large language model code generation mostly targets general benchmarks such as HumanEval, MBPP, and CodeContests {[}7{]}{[}8{]}{[}9{]}, with the core goal of measuring whether a model can generate correct code. The problems in university programming examinations are different: teachers are not concerned only with whether a model can solve a problem, but more with the relative difficulty of a problem within their own course's student population, whether the structure of an entire exam is reasonable, whether exams across different parallel classes are comparable, and whether there is a low-cost, reviewable external reference at the item-setting stage. Therefore, this paper shifts the research perspective from ``model capability evaluation'' to ``course examination evaluation,'' positioning the data produced by AI as auxiliary evidence in the interpretation of problem and exam difficulty.

This paper conducts research based on real course examination data, but its purpose is not to simply showcase the processing results of several datasets; rather, it unfolds layer by layer around validity evidence for the AI difficulty ruler. The first type of data comes from a synchronous computer-based examination experiment: 10 large language models and students answered the same exam problem set, with all code judged via OJ feedback, used to test whether AI answering responses can form a relative problem-difficulty ordering consistent with student performance. The second type of data comes from the final computer-based examinations of 11 parallel classes in the same semester: a total of 79 programming problems, with each exam having problem statements, reference answers, and class-wide whole-exam score distributions, used for subsequent testing of whether low-cost AI review can enter parallel-class exam comparison. The third type of data comes from 26 problems across 4 final computer-based examinations of the same teacher's Data Structures and Algorithms B, used to observe whether the same ruler can support cross-semester problem-risk tracking. The fourth type of data comes from 106 problems across 16 past final computer-based examinations of the same teacher's Introduction to Computing B, used to test the applicability boundary of this ruler in an introductory course and low-difficulty problems. The latter three types of data have all completed a reproducible local pipeline, auditable API review batches, content auditing, human review, and statistical aggregation, and their review evidence has a third-party endpoint single reviewer as its source boundary. Together, the four types of data point to a core question: under what conditions can AI-generated problem-side evidence support the difficulty interpretation of real course examinations, and under what conditions must it fall back to a descriptive reference.

Accordingly, this paper poses the following research questions:

RQ1: In a real online judge environment, can synchronous answering data from large language models form a relative difficulty ordering consistent with students' problem pass rates?

RQ2: In a larger-scale parallel-class scenario, can auditable API structured review based on problem statements and reference answers reflect student-population performance at the exam level and problem level?

RQ3: How do item-bank exposure, item-count pressure, non-attempt rate, and error-type distribution help explain the applicability boundary of the AI difficulty ruler?

RQ4: In an item-setting scenario spanning semesters and course levels for the same teacher, can the auditable API review ruler support longitudinal comparison of exam difficulty and problem risk?

\hypertarget{theoretical-foundations-and-analytical-framework}{%
\section{Theoretical Foundations and Analytical Framework}\label{theoretical-foundations-and-analytical-framework}}

\hypertarget{the-educational-evaluation-reform-perspective-ai-as-auxiliary-evidence}{%
\subsection{The Educational Evaluation Reform Perspective: AI as Auxiliary Evidence}\label{the-educational-evaluation-reform-perspective-ai-as-auxiliary-evidence}}

As stated above, educational evaluation reform policy encourages the use of modern information technologies such as artificial intelligence and big data to innovate evaluation tools {[}5{]}{[}6{]}. Programming examinations naturally have a technical foundation that is submittable, judgeable, and recordable, making them suitable for exploring intelligent auxiliary evaluation methods. However, this paper does not treat AI as the agent that automatically evaluates students; instead, it positions AI answering and AI review results as auxiliary evidence in item-setting review and exam-structure interpretation.

This positioning has two implications. First, the AI ruler is oriented toward problems and exams, not toward individual students. It answers ``what difficulty structure does this problem or this exam present to several code-generation systems,'' rather than ``what evaluation should a particular student receive.'' Second, AI evidence needs to be interpreted jointly with student-population performance, problem provenance, the teacher's preset objectives, and the course's training path; it cannot be directly converted into score-adjustment rules apart from the instructional context.

From the perspective of educational evaluation theory, the key question for an evaluation tool is not merely whether it can produce a score, but whether the interpretation and use of the score are adequately supported by evidence. Messick's validity theory emphasizes that validity is an integrated judgment formed around the interpretation and use of test scores, encompassing both evidence from content, structure, and external relations, and consideration of evaluation consequences {[}10{]}. Kane further organizes validity examination into an argument chain centered on score interpretation and use, emphasizing link-by-link scrutiny of the interpretive chain {[}11{]}{[}12{]}. The ``Standards for Educational and Psychological Testing'' likewise treats validity, reliability, fairness, and responsibility for test use as basic requirements for building evaluation tools {[}13{]}. Accordingly, this paper does not understand AI scoring as an independent ``difficulty ground truth,'' but places it within an evidence chain jointly constituted by student performance, online judge process data, teachers' professional judgment, and problem-context information, examining whether it can support limited and clearly defined evaluation uses.

\hypertarget{the-educational-measurement-perspective-from-a-single-pass-rate-to-multi-source-evidence}{%
\subsection{The Educational Measurement Perspective: From a Single Pass Rate to Multi-Source Evidence}\label{the-educational-measurement-perspective-from-a-single-pass-rate-to-multi-source-evidence}}

Item response theory (IRT) suggests to us that problem difficulty can be estimated through answering responses {[}14{]}{[}15{]}; but in real course scenarios, programming-problem difficulty is not manifested only as pass rate. The modeling ability, algorithm selection, state maintenance, boundary handling, debugging experience, and time allocation required for a student to complete a problem all affect the actual difficulty of the problem. The pass status, number of attempts, running time, and feedback types provided by online judge platforms enable researchers to understand problem difficulty from both an outcome and a process angle.

This paper draws on the idea in educational measurement of ``estimating problem difficulty from answering responses,'' but does not pursue strict IRT parameter interpretation in small samples. The 10 models and 8 problems in the solving-based experiment are suitable for verifying ordering consistency and methodological feasibility; the 11 exams and 79 problems in the cross-sectional review-based calibration, the 4 examinations and 26 problems in the Data Structures and Algorithms B longitudinal comparison, and the 16 examinations and 106 problems in the Introduction to Computing B sample are used to test whether the auditable API review workflow can expand the coverage of problems and exams. The three sets of structured scores used in the main text all come from the same batch of auditable API review (model identifier gpt-5.6-sol, invoked via a third-party OpenAI-compatible endpoint), which preserves the endpoint, temperature, prompt/schema hashes, raw responses, and run metadata, and uses student performance, item-bank exposure, and course level as external references.

At the same time, the digital process data of programming examinations also connects this paper with learning analytics research. Learning analytics emphasizes using data generated by learners to understand and optimize the learning process {[}16{]}, but related research also reminds researchers that they must attend to the contextuality of data interpretation, the value orientation of indicator selection, and the action consequences of evaluation feedback {[}17{]}{[}18{]}. This paper's use of pass rate, non-attempt rate, number of submissions, and error type is not intended to build an automatic profile of individual students, but to extend problem-difficulty interpretation from a single outcome indicator to an evidence structure combining ``outcome--process--mechanism.'' This approach is consistent with the ``evidence-centered'' tradition of evaluation design: evaluation conclusions should be built on a defensible evidence chain, not on a single score {[}19{]}{[}20{]}.

\hypertarget{two-forms-of-the-ai-ruler}{%
\subsection{Two Forms of the AI Ruler}\label{two-forms-of-the-ai-ruler}}

This paper distinguishes two forms of the AI ruler. The first is the solving-based AI ruler: the AI reads the problem statement, generates code, submits it to the OJ, and performs a limited number of revision rounds based on judging feedback. Its advantage is that it is close to the real examination process and can produce process data such as pass rate, number of attempts, running time, and error type; its shortcoming is that it has relatively high running costs and is affected by model access, platform status, and feedback-repair strategy. The second is the review-based AI ruler: the AI does not submit code but, based on the problem statement, constraints, and reference answer, provides structured scores for concept difficulty, implementation difficulty, debugging difficulty, complexity risk, statement-reading difficulty, and overall difficulty. Its potential advantage is that it is low-cost and easy to apply across more exams, making it suitable for parallel-class item-setting review and cross-sectional exam comparison; its shortcoming is that it lacks real judging feedback, its scoring results need to be validated by student performance, teacher judgment, or multi-reviewer agreement, and it must preserve reviewable source metadata such as model version, prompt hash, run date, endpoint, and sampling parameters.

This paper uses the solving-based evidence of Study One as the precondition for the review-based extension: if the problem-difficulty ordering obtained from multiple models answering synchronously is highly consistent with student pass rates, then AI review can be used as a low-cost extension method for larger-scale exam-structure analysis. But such an extension remains exploratory research and cannot be equated with an already-stably-calibrated measurement model.

\hypertarget{intrinsic-difficulty-and-effective-exam-difficulty-the-dual-role-of-the-exposure-variable}{%
\subsection{Intrinsic Difficulty and Effective Exam Difficulty: The Dual Role of the Exposure Variable}\label{intrinsic-difficulty-and-effective-exam-difficulty-the-dual-role-of-the-exposure-variable}}

Programming problems have intrinsic difficulty, as well as an effective difficulty in the examination context. Intrinsic difficulty is mainly determined by algorithmic concepts, implementation path, boundary conditions, complexity risk, and statement-reading burden; effective exam difficulty is also affected by whether the problem was exposed in advance, whether it is an original item from an item bank, whether students trained intensively on similar problems, exam duration, item count, and score structure. The same problem is not equivalent for a student who has not seen the problem statement and one who has already practiced it in an item bank.

Therefore, in addition to review-based AI difficulty, this paper introduces the problem-exposure variable. The raw AI score is used to describe the intrinsic difficulty of a problem; the exposure-adjusted difficulty is used to describe the effective difficulty in the examination context. Neither is directly used for score adjustment; rather, they serve parallel-class exam interpretation.

The second role of the exposure variable is that the AI reviewer reads only the problem statement and reference answer and does not ``know'' which problems are original items from the item bank. This means that the deviation between AI difficulty and student performance itself constitutes an interpretable signal. When a problem receives a high AI score but students' pass rate is also high, it may indicate an item-bank exposure effect; when the AI score is moderate but students' pass rate is very low, it may indicate item-count pressure, insufficient time, or insufficient instructional coverage. In other words, the exposure variable is not only a passive adjustment term; it is itself an analytical output of the AI ruler---AI helps identify which exams have had their effective difficulty systematically altered by non-problem factors.

\hypertarget{the-humanai-collaboration-perspective-serving-the-teachers-item-setting-community}{%
\subsection{The Human--AI Collaboration Perspective: Serving the Teachers' Item-Setting Community}\label{the-humanai-collaboration-perspective-serving-the-teachers-item-setting-community}}

Large language models are already able to complete a large number of routine programming problems, which changes the frame of reference for programming course examinations. If a mainstream model can quickly pass a problem, that problem may be closer to basic implementation or template-based training; if multiple models fail, repeatedly repair, or show high running risk on the same problem, that problem may have a higher threshold of modeling, complexity, or boundary handling. But model performance itself is also affected by prompt, model version, training-data exposure, submission language, and feedback-repair strategy, and cannot be regarded as a stable, objective arbiter.

Therefore, this paper adopts a human--AI collaboration framework: AI provides reproducible external difficulty evidence, the online judge platform provides objective judging feedback, and teachers interpret it in combination with course objectives and student background. This framework emphasizes ``AI augmenting teacher judgment'' rather than ``AI replacing teacher judgment.''

This positioning also responds to the governance issues that arise once generative AI enters educational evaluation. Existing research points out that generative AI can support feedback and evaluation of complex learning performance, but its outputs require transparent boundaries of use, human review, and accountability {[}21{]}{[}22{]}{[}23{]}. Recent studies in the ``Peking University Education Review'' on university digital transformation, reasoning language models, human-centered teaching philosophy, learning ethics, and comprehensive assessment frameworks also suggest that AI educational applications are not merely a matter of tool efficiency, but also involve digital educational infrastructure, process diagnosis, learning ethics, expansion of evaluation goals, and human-centered responsibility {[}24{]}{[}25{]}{[}26{]}{[}27{]}{[}28{]}. In the field of computing education, AI code generation has been regarded as a structural change bearing both opportunities and challenges {[}29{]}; researchers have observed that programming teachers' attitudes toward AI are shifting from ``resistance'' to ``acceptance and adjusting teaching'' {[}30{]}, and early Codex experiments also suggested that beginners may over-rely on code generation and thereby weaken basic skills {[}31{]}{[}32{]}. Students' and teachers' use of generative AI simultaneously involves expectations of efficiency, reconstruction of learning goals, and concerns about academic integrity {[}33{]}{[}34{]}. Therefore, this paper designs the AI difficulty ruler as discussion material for the teachers' item-setting community, rather than an automatic decision system; its reasonable use is to help teachers discover problem risks, compare exam structures, and reflect on evaluation goals, rather than to replace teachers' professional judgment about course objectives and student development.

\hypertarget{research-design-and-data-sources}{%
\section{Research Design and Data Sources}\label{research-design-and-data-sources}}

\hypertarget{overall-design-organizing-data-around-validity-evidence}{%
\subsection{Overall Design: Organizing Data Around Validity Evidence}\label{overall-design-organizing-data-around-validity-evidence}}

This paper adopts a progressive research design rather than treating AI scoring directly as the ground truth of problem difficulty. The first step establishes baseline validity evidence through solving-based calibration: 10 large language models and students complete the same 8 programming problems in the same OJ exam environment, and real judging feedback is used to form process-level indicators such as AI pass rate, number of attempts, and running time, in order to test whether these indicators can form a consistent ordering with the students' problem pass rate. The second step expands coverage through review-based calibration: the study objects are extended to 79 programming problems from the final machine exams of 11 parallel classes, using a single unified reviewer (whose single-reviewer design derives from a historical selection basis in the first-stage ChatGPT solving performance, and which actually runs as an auditable API batch on a third-party endpoint), which outputs structured difficulty scores based on the problem statement and reference solution, and which is compared against student-group performance at the exam level and the problem level. The third step introduces longitudinal and cross-course boundary tests: reusing the same review pipeline, it processes 4 final machine exams of the same teacher's Data Structures and Algorithms B in Spring 2024, Spring 2025, Fall 2025, and Spring 2026, as well as 16 final machine exams of Introduction to Computing B from Fall 2008 to Fall 2025, observing the explanatory value of the AI ruler in cross-semester tracking of problem-setting quality and in cross-course transfer. The problem extraction, student statistics, input construction, and the three groups of auditable API reviews for the second and third steps have all been completed; the source boundary of the review evidence is disclosed in Section 3.5.

This design serves three levels of the validity argument. First, the solving-based experiment answers whether AI can produce a problem ordering consistent with student performance in a real judging environment; second, the horizontal parallel-class sample answers whether a low-cost review-based ruler can be extended to more exams and obtain external criteria through student pass rate, non-attempt rate, and error types; third, the longitudinal sample answers the applicable boundary of the same ruler across different course levels. The specific technical workflow includes extraction of problems and reference solutions, construction of AI solving or structured review inputs, aggregate statistics from OJ ranking/status pages, and merged analysis of AI difficulty indicators with student performance. All student data are processed into group-aggregate statistics, retaining no personally identifiable information.

\hypertarget{first-stage-sample-8-problem-synchronous-machine-exam}{%
\subsection{First-Stage Sample: 8-Problem Synchronous Machine Exam}\label{first-stage-sample-8-problem-synchronous-machine-exam}}

The first-stage exam contains 8 programming problems, covering array processing, sorting, simulation, graph theory, dynamic programming, data-structure maintenance, and combinatorial counting. On the student side, the 120 students who actually took the exam are the statistical objects; an additional 2 students did not take the final machine exam for personal reasons and are not included in the pass rate, AC-count distribution, or score statistics. On the AI side, there are 10 representative large language models, covering web-end models, API models, and locally deployed models. The models read the problem statements, generate code, and submit to the OJ through automated scripts, and perform a limited number of repair rounds based on non-AC feedback.

\hypertarget{horizontal-sample-11-parallel-class-exams-and-79-problems}{%
\subsection{Horizontal Sample: 11 Parallel-Class Exams and 79 Problems}\label{horizontal-sample-11-parallel-class-exams-and-79-problems}}

The horizontal sample comes from the final machine exams of 11 parallel classes of the same course. The number of problems per exam ranges from 5 to 13, the student-group statistics cover 28 to 120 actual participants, and the 11 exams total 1031 person-times. Table 1 presents the basic information of the 11 exams. 8 exams use the solved-count distribution as the student performance indicator, and 3 exams use the score distribution as the student performance indicator.

\textbf{Table 1 Basic information of the 11 parallel-class final machine exams}

\begin{longtable}[]{@{}
  >{\raggedright\arraybackslash}p{(\columnwidth - 10\tabcolsep) * \real{0.1364}}
  >{\raggedleft\arraybackslash}p{(\columnwidth - 10\tabcolsep) * \real{0.1818}}
  >{\raggedleft\arraybackslash}p{(\columnwidth - 10\tabcolsep) * \real{0.1818}}
  >{\raggedright\arraybackslash}p{(\columnwidth - 10\tabcolsep) * \real{0.1364}}
  >{\raggedleft\arraybackslash}p{(\columnwidth - 10\tabcolsep) * \real{0.1818}}
  >{\raggedleft\arraybackslash}p{(\columnwidth - 10\tabcolsep) * \real{0.1818}}@{}}
\toprule\noalign{}
\begin{minipage}[b]{\linewidth}\raggedright
Exam
\end{minipage} & \begin{minipage}[b]{\linewidth}\raggedleft
Items
\end{minipage} & \begin{minipage}[b]{\linewidth}\raggedleft
Actual participants
\end{minipage} & \begin{minipage}[b]{\linewidth}\raggedright
Student metric
\end{minipage} & \begin{minipage}[b]{\linewidth}\raggedleft
Exposed items
\end{minipage} & \begin{minipage}[b]{\linewidth}\raggedleft
Exposure ratio
\end{minipage} \\
\midrule\noalign{}
\endhead
\bottomrule\noalign{}
\endlastfoot
Class 4 & 8 & 115 & Score & 0 & 0.000 \\
Class 7 & 6 & 28 & Solved count & 0 & 0.000 \\
Class 8 & 8 & 61 & Solved count & 8 & 1.000 \\
Class 9 & 13 & 93 & Solved count & 3 & 0.231 \\
Class 10 & 5 & 107 & Solved count & 0 & 0.000 \\
Class 12 & 6 & 107 & Solved count & 0 & 0.000 \\
Class 13 & 5 & 94 & Score & 0 & 0.000 \\
Class 14 & 7 & 95 & Solved count & 0 & 0.000 \\
Class 15 & 8 & 120 & Solved count & 0 & 0.000 \\
Class 16 & 7 & 107 & Solved count & 0 & 0.000 \\
Class 2 & 6 & 104 & Score & 1 & 0.167 \\
\end{longtable}

Among them, all 8 problems of Class 8 are original problems from the OJ practice item bank; Class 9's 02816 ``Red and Black,'' 24729 ``Bracket-Nested Tree,'' 05455 ``Level-Order Traversal of a Binary Search Tree,'' and Class 2's 29803 ``Crossing the Fire Line'' are also original problems from the item bank. This paper labels the above 12 problems as item-bank-exposed problems.

\hypertarget{ai-solving-based-difficulty-indicator}{%
\subsection{AI Solving-Based Difficulty Indicator}\label{ai-solving-based-difficulty-indicator}}

For a first-stage problem p, let the number of participating AI models be M, and let whether model j passes be recorded as \(AC_{p,j}\); for the models that pass the problem, let their average number of attempts be \(\overline{Attempts}_p\), and the mean of the ratio of the AC code's running time to the problem's time limit \(TL_p\) be \(\overline{Time}_p/TL_p\). This paper constructs the difficulty indicator:

\[
d_p = \alpha \left(1 - \frac{\sum_{j=1}^{M} AC_{p,j}}{M}\right)
 + \beta \min\left(1, \frac{\overline{Attempts}_p}{5}\right)
 + \gamma \min\left(1, \frac{\overline{Time}_p}{TL_p}\right)
\]

where α=0.6, β=0.2, γ=0.2. The weights are based on the following considerations: the pass rate is the most direct proxy indicator of difficulty and is assigned the highest weight; the number of attempts reflects the process cost of solving, and the running-time ratio reflects computational-efficiency pressure, and each of the latter two takes a smaller weight. The sensitivity analysis of the attempt-count truncation value of 5 (with truncation values of 3, 5, 10, and no truncation) shows that across the 8 problems the maximum number of model attempts is 3.00, and under each truncation scheme the Spearman ρ between \(d_p\) and the student pass rate is -0.905, with completely consistent ordering. This indicator, dominated by the AI non-pass proportion and supplemented by the number of attempts and the time cost, is used to form a relative difficulty ordering of the problems.

\hypertarget{ai-review-based-difficulty-indicator}{%
\subsection{AI Review-Based Difficulty Indicator}\label{ai-review-based-difficulty-indicator}}

The horizontal and longitudinal review-based calibration adopts a single-reviewer design. The historical selection basis for this design comes from the first-stage solving experiment: among the 10 participating models, ChatGPT was the only model that passed all 8 problems (AK) and cracked the extremely difficult problem I30547, and therefore its subsequent version was selected as the candidate for the unified reviewer. But two points must be made clear. First, a model's strong problem-solving ability does not equal strong scoring ability---solving a problem (generating a correct answer) and judging a problem's difficulty (a metacognitive estimate of students' difficulty) are two different cognitive operations. Second, the review batch actually used in this paper is invoked through a third-party OpenAI-compatible endpoint, and the model identifier in both the request and the endpoint's return is gpt-5.6-sol; this label cannot certify that the upstream is an official OpenAI model, so this paper does not describe this batch as ``official ChatGPT review.''

The three review batches (79 problems, 26 problems, 106 problems) were run on July 10, 2026 under a unified protocol: \texttt{temperature=0}, \texttt{reasoning\_effort=high}, and JSON-schema-constrained output; the input for each problem includes the problem statement, input/output format, samples, constraints, reference solution, course identifier, course level, and target student group; the output is accompanied by the model identifier, invocation date, endpoint, temperature, the SHA-256 of the prompt and schema, the raw API response, and run metadata, and all 211 reviews were strictly validated by \texttt{validate\_ai\_reviews.py\ -\/-require-provenance} (cross-consistency of the normalized output, raw response, and run metadata with the hashes of the input, prompt, schema, and runner). The content audit found two categories of text-level issues and discloses them as review items: for individual problems, the complexity description of the reference solution was filled in according to the ideal solution rather than the given reference code; and for repeated problems, there are two formulations of whether the cost of big-integer arithmetic is counted into the bit length. These issues involve only free-text fields and did not enter any numerical analysis. The audit also confirmed that the estimated student pass-rate interval of the review output corresponds highly mechanically to the overall difficulty, so this paper does not treat the two as mutually independent AI evidence. Among the 106-problem sample, of the 13 groups of repeated inputs with identical problem statements and reference code, 5 groups differ by 1 level in overall difficulty, and this paper reports the corresponding sensitivity test in Section 6; \texttt{temperature=0} does not constitute a guarantee of output determinism.

The input for each problem includes the problem statement, input/output format, samples, constraints, and reference solution. The language of the reference solution is faithfully recorded according to the source document; in the horizontal 79-problem sample, 78 problems are Python and 1 problem is C++. The reviewer outputs the following dimensions on a 1-to-5 scale, where a higher score indicates higher difficulty or risk:

\begin{itemize}
\tightlist
\item
  Concept difficulty: the requirements of algorithmic ideas, data structures, and mathematical modeling.
\item
  Implementation difficulty: the complexity of converting ideas into code.
\item
  Debugging difficulty: the difficulty of boundary conditions, state maintenance, input/output, and error localization.
\item
  Complexity risk: the risk that students adopt a naive algorithm leading to TLE/MLE or complexity mismatch.
\item
  Reading difficulty: the difficulty of understanding the problem statement, extracting conditions, and modeling the problem.
\item
  Overall difficulty: a comprehensive assessment.
\end{itemize}

At the exam level, the mean of overall difficulty, the number of high-difficulty problems, and the proportion of high-difficulty problems are computed. This paper defines a problem with overall difficulty greater than or equal to 4 as a high-difficulty problem. The single-reviewer result can only be interpreted as a preliminary/exploratory ruler and cannot replace a multi-reviewer consistency test. Because the same reviewer may exhibit output variation across invocations, and model-version upgrades may affect review consistency, the review-based result must be regarded as a verification of methodological feasibility, rather than an already stably calibrated measurement model {[}34{]}. As an archival-level sensitivity comparison, this study compares, in three columns, 12 problem IDs from this API batch with the Opus 4.8 label archive and the DeepSeek label archive carried over in the repository: the exploratory ICC(C,1) of the overall difficulty across the three columns is 0.9397, the pairwise Spearman correlations range between 0.928 and 0.930, and the range of overall difficulty of the 12 problems does not exceed 1 level. However, the two old columns have neither raw responses, run metadata, problem-version binding, nor prompt binding, their run protocols are not comparable, and the DeepSeek identity further relies on configuration records outside the author's repository; moreover, the 12 problems are a non-random purposive sample covering 6 exams and covering difficulty 1-5 under the old scoring, and widening the difficulty spectrum may inflate the correlation and cannot be extrapolated to all 79 problems. Therefore, this comparison is reported only as archival audit material and cannot be written up as a formal three-reviewer reliability or an official cross-developer validation.

In addition to AI structured scoring, this paper also incorporates the course teacher's experiential manual review as an explanatory cross-check. The first author is the teacher of Class 15 of ``Data Structures and Algorithms B'' in Spring 2026, directly participating in that class's final-machine-exam problem setting, post-exam score interpretation, and review of comparable exams over the years, and possesses insider expert knowledge of the problems' preset levels, course training coverage, students' common errors, and problem-setting goals. This paper does not treat this teacher experience as independent blind-review data, nor does it compute manual-review consistency on this basis; its role is to help explain the consistency or deviation between AI scoring and student performance, especially by providing course-internal evidence on contextual variables such as problem-quantity pressure, item-bank exposure, high-difficulty-problem proportion, and complexity risk. In other words, the manual review in this paper belongs to ``explanatory validation with the participation of the teacher's professional judgment,'' rather than an independent external expert review.

\hypertarget{item-bank-exposure-adjustment}{%
\subsection{Item-Bank Exposure Adjustment}\label{item-bank-exposure-adjustment}}

For a problem p, let the AI raw overall difficulty be \(d^{raw}_p\) and the exposure discount be \(discount_p\); then the effective difficulty is:

\[
d^{adj}_p = d^{raw}_p \times (1 - discount_p)
\]

For the 12 practice item-bank original problems, this paper adopts a conservative discount discount\_p=0.25; for the remaining problems not marked as exposed, 0 is taken for now. The choice of the discount coefficient requires methodological justification. 0.25 is a conservative setting---it assumes that item-bank visibility reduces the effective difficulty of a problem by one quarter, but does not over-adjust it. This paper gives the design of the exposure-discount sensitivity analysis in Section 3.9.

\hypertarget{student-group-performance-indicators}{%
\subsection{Student-Group Performance Indicators}\label{student-group-performance-indicators}}

The student performance in the horizontal review-based calibration comes from the whole-exam score distribution of each exam. For the solved-count distribution, the average solved count, median solved count, full-completion rate, low-performance rate, and normalized average performance are computed, where the normalized average performance is the average solved count divided by the exam's declared item count; the low-performance rate is defined as the proportion of students whose solved count is less than or equal to 1 problem. For the score distribution, the normalized average performance is the average score divided by the exam's full score (computed as the declared item count multiplied by 100 points per problem); the low-score rate is defined as the proportion of students whose score is lower than or equal to 40\% of the full score. Because different classes adopt the two statistical calibers of solved count and score, this paper simultaneously reports, in the relevant analysis, the results of all 11 exams and the results of the 8 exams that contain only solved-count distributions.

To further advance from the exam level to the problem level, this paper also scrapes the OJ ranking pages and status pages of the 11 exams, forming a problem-level student behavior table of 79 problems. The ranking page is used to compute, for each problem, the actual-participant denominator, the number of passers, the pass rate, the number of attempters, the number of non-attempters, the non-attempt rate, the number of submissions visible in the ranking cells, and the distribution of the time to first AC under the ACM/ICPC ranking rule; the status page is used to count the distribution of submission results for each problem, including Accepted, Wrong Answer, Time Limit Exceeded, Memory Limit Exceeded, Runtime Error, Compile Error, Presentation Error, Output Limit Exceeded, and Waiting. The time to first AC is aggregated into a problem-level distribution by 0-30 minutes, 30-60 minutes, 60-90 minutes, 90-120 minutes, 120-180 minutes, and more than 180 minutes. To protect student privacy, the scraping process uses user identifiers only in memory for deduplication and filtering, and the output table retains only problem-level aggregate statistics, saving no names, accounts, submission IDs, individual times to first AC, or code.

Problem-level analysis takes ``one problem in one exam'' as the unit, for a total of 79 observations. The pass rate reflects the final completion result, the non-attempt rate reflects students' give-up behavior under time pressure and difficulty judgment, and the status submission-result distribution reflects the error-type structure.

\hypertarget{longitudinal-sample-and-boundary-test}{%
\subsection{Longitudinal Sample and Boundary Test}\label{longitudinal-sample-and-boundary-test}}

To test whether the AI ruler can be used for the same teacher's cross-semester tracking of problem-setting quality, this paper selects the final-machine-exam problems of Data Structures and Algorithms B taught by the first author over the years as the first group of longitudinal sample. This sample covers 4 final machine exams in Spring 2024, Spring 2025, Fall 2025, and Spring 2026, for a total of 26 problems; among them, the 8 problems of Spring 2026 reuse the Class 15 data from the first-stage synchronous solving experiment and the parallel-class extension, and the remaining 18 problems from 3 exams are extracted---problem statements, links, and reference code---from the problem documents over the years. The longitudinal sample maintains the same scoring schema and problem-level student behavior indicators as the horizontal review-based calibration, so as to observe the descriptive trend of the same teacher's problem-setting structure changing over semesters. This sample does not introduce the item-bank exposure discount, avoiding mixing cross-semester structural changes with the exposure parameter in the interpretation.

The second group of longitudinal sample is the final-machine-exam problems of Introduction to Computing B taught by the first author over the years. This sample covers 16 final machine exams: Fall 2008, Fall 2013, Fall 2014, Fall 2016, Fall 2017, Fall 2018, Fall 2019, Fall 2020 non-international students, Fall 2020 international students, Fall 2021 non-international students, Fall 2021 international students, Fall 2022 non-international students, Fall 2022 international students, Fall 2023, Fall 2024, and Fall 2025, for a total of 106 problems. The function of this sample is not simply to increase the number of problems, but to test the discriminative ability of the AI review ruler in an introductory course, low-difficulty problems, and a longer historical span. If an international-student exam in the document marks a repeated problem with ``same as before,'' the extraction program reuses the problem statement and reference code from the previous occurrence with the same problem number and title. The OJ ranking pages and status pages of all 106 problems were successfully parsed, forming problem-level student pass rate, non-attempt rate, number of submissions, time-to-first-AC distribution, and error-type statistics.

\hypertarget{exposure-discount-sensitivity-analysis-design}{%
\subsection{Exposure-Discount Sensitivity Analysis Design}\label{exposure-discount-sensitivity-analysis-design}}

To test whether the influence of the exposure adjustment on the correlation between AI difficulty and student performance depends on a particular discount choice, this paper conducts a four-level sensitivity analysis on the discount coefficient \(discount_p\) ∈ \{0, 0.10, 0.25, 0.40\}. The four levels cover the range from ``no adjustment'' to ``strong adjustment'': 0 corresponds to completely ignoring the exposure effect, 0.10 corresponds to a light discount, 0.25 is the conservative value adopted in this paper's main analysis, and 0.40 corresponds to a stronger discount. For each discount level, this paper recomputes the Pearson and Spearman correlation coefficients at the problem level (all 79 problems, only the 67 non-exposed problems) and at the exam level (all 11 exams, only the 8 exams under the solved-count caliber).

\hypertarget{stage-one-results-validity-of-solving-based-calibration}{%
\section{Stage-One Results: Validity of Solving-Based Calibration}\label{stage-one-results-validity-of-solving-based-calibration}}

\hypertarget{ai-solving-produces-a-clear-problem-gradient}{%
\subsection{AI Solving Produces a Clear Problem Gradient}\label{ai-solving-produces-a-clear-problem-gradient}}

The solving performance of 10 AI models on 8 problems formed a clear gradient. Basic problems and routine medium problems were passed by all or nearly all models; medium-high-difficulty problems began to differentiate models; the extremely hard problem was passed by only 1 model. In terms of specific model performance, ChatGPT was the only model that solved all 8 problems (AK) and the only model that passed the extremely hard problem I30547. Table 2 reports AI solving performance and difficulty indices.

\textbf{Table 2 AI solving performance, difficulty indices, and student pass rates in the synchronous 8-problem machine exam}

\begin{longtable}[]{@{}
  >{\raggedright\arraybackslash}p{(\columnwidth - 12\tabcolsep) * \real{0.1154}}
  >{\raggedleft\arraybackslash}p{(\columnwidth - 12\tabcolsep) * \real{0.1538}}
  >{\raggedleft\arraybackslash}p{(\columnwidth - 12\tabcolsep) * \real{0.1538}}
  >{\raggedleft\arraybackslash}p{(\columnwidth - 12\tabcolsep) * \real{0.1538}}
  >{\raggedleft\arraybackslash}p{(\columnwidth - 12\tabcolsep) * \real{0.1538}}
  >{\raggedleft\arraybackslash}p{(\columnwidth - 12\tabcolsep) * \real{0.1538}}
  >{\raggedright\arraybackslash}p{(\columnwidth - 12\tabcolsep) * \real{0.1154}}@{}}
\toprule\noalign{}
\begin{minipage}[b]{\linewidth}\raggedright
Problem
\end{minipage} & \begin{minipage}[b]{\linewidth}\raggedleft
AI pass rate
\end{minipage} & \begin{minipage}[b]{\linewidth}\raggedleft
Mean attempts of AC models
\end{minipage} & \begin{minipage}[b]{\linewidth}\raggedleft
Time-to-limit ratio
\end{minipage} & \begin{minipage}[b]{\linewidth}\raggedleft
Difficulty index \(d_p\)
\end{minipage} & \begin{minipage}[b]{\linewidth}\raggedleft
Student pass rate
\end{minipage} & \begin{minipage}[b]{\linewidth}\raggedright
Teacher-intended level
\end{minipage} \\
\midrule\noalign{}
\endhead
\bottomrule\noalign{}
\endlastfoot
E30646 & 100\% & 1.00 & 0.020 & 0.044 & 95.8\% & Basic \\
E30930 & 100\% & 1.00 & 0.265 & 0.093 & 88.3\% & Basic \\
M30680 & 100\% & 1.00 & 0.023 & 0.045 & 79.2\% & Medium \\
M30874 & 100\% & 1.00 & 0.083 & 0.057 & 83.3\% & Medium \\
T30913 & 100\% & 1.20 & 0.360 & 0.120 & 25.0\% & Harder \\
M30947 & 90\% & 1.89 & 0.202 & 0.176 & 5.0\% & Medium-hard \\
U30919 & 90\% & 1.00 & 0.863 & 0.273 & 1.7\% & Hard \\
I30547 & 10\% & 3.00 & 0.110 & 0.682 & 0.0\% & Extremely hard \\
\end{longtable}

Note: Student pass rates are computed on the OJ ranking-page basis, with the denominator being the 120 enrolled participants of that machine exam (2 external participants excluded); values are taken from \texttt{paper\_tables/problem\_student\_stats.csv}. An earlier version of this paper used, for this column, a different summary whose basis was not archived, and for a few problems the number of passing students differs by ±1 from the current basis; under both bases the ranking of the 8 problems' pass rates is identical, and the rank-correlation results in Section 4.2 are unaffected.

Viewed by the \(d_p\) ordering, the problems form a continuous ``basic-medium-medium-high-extremely-hard'' gradient, indicating that the AI solving process can capture the main levels of teacher-intended difficulty. I30547 was passed only by ChatGPT and had the highest composite difficulty index; although U30919 had a 90\% AI pass rate, its running time was close to the time limit, so its composite difficulty is higher than that of ordinary medium problems. This shows that the number of attempts and time cost can compensate for the completion cost that a single pass rate cannot express. ChatGPT's full-solve performance in Stage One also constitutes the empirical basis for choosing a unified reviewer in the subsequent review-based extension.

\hypertarget{ai-indices-show-significant-rank-correlation-with-student-performance}{%
\subsection{AI Indices Show Significant Rank Correlation with Student Performance}\label{ai-indices-show-significant-rank-correlation-with-student-performance}}

The Stage-One results show that the Spearman rank correlation between AI pass rate and student problem pass rate is 0.866, with an exact two-sided permutation p=0.0119; the Spearman rank correlation between the composite difficulty index \(d_p\) and student problem pass rate is -0.905, with an exact two-sided permutation p=0.0046. The direction of correlation matches expectations: the higher the AI pass rate, the higher the student pass rate; the higher the AI composite difficulty, the lower the student pass rate. The composite index exhibits stronger ranking consistency than pass rate alone, indicating that the number of attempts and time cost during solving carry additional explanatory value.

At the same time, there are absolute-level differences between AI and student performance. On medium-high-difficulty problems such as T30913, M30947, and U30919, the AI pass rate is markedly higher than the student pass rate. This shows that once a large language model forms a correct algorithmic idea, the cost of generating runnable code may be lower than for students; whereas students are additionally affected by coding proficiency, debugging experience, exam time allocation, and psychological pressure. Therefore, the AI scale is suitable for relative difficulty ranking and problem-setting review, but not for directly predicting the proportion of students who pass.

\hypertarget{robustness-check}{%
\subsection{Robustness Check}\label{robustness-check}}

To test whether the \(d_p\) ordering depends on a single weight setting, the study compared 4 weight schemes: pass-rate-first (0.6/0.2/0.2), balanced (0.5/0.25/0.25), pass-rate-emphasized (0.7/0.15/0.15), and process-indicator-emphasized (0.4/0.3/0.3). Under all 4 schemes, I30547 is always the hardest problem, and U30919 and M30947 are always in the high-difficulty region. An exploratory Rasch fit (weak-prior MAP estimation) was also performed on the 8 problems: the 5 fully-solved problems have no finite maximum-likelihood difficulty estimate and are tied at the lower end of the scale, M30947 and U30919 cannot be distinguished because of identical response patterns, and only I30547 obtains a clear high-difficulty location; the direction of the difficulty ordering agrees with \(d_p\) and student pass rate, but the parameter values are unstable and are used only for directional observation, not as independent evidence of IRT validity. This result provides an antecedent basis for the subsequent adoption of AI for a larger-scale review-based extension.

\hypertarget{cross-sectional-review-based-calibration-results-ai-review-difficulty-and-parallel-class-student-statistics}{%
\section{Cross-Sectional Review-Based Calibration Results: AI Review Difficulty and Parallel-Class Student Statistics}\label{cross-sectional-review-based-calibration-results-ai-review-difficulty-and-parallel-class-student-statistics}}

\begin{quote}
\textbf{Provenance boundary note:} The ``AI review difficulty'' in this section and in Sections Six and Seven all comes from the auditable API review batch of July 10, 2026 (a third-party OpenAI-compatible endpoint, with both the request and returned model identifier being gpt-5.6-sol). This label cannot certify that the upstream is an official OpenAI model, and the review evidence should be understood as a single-reviewer exploratory scale with an auditable provenance.
\end{quote}

\hypertarget{exam-differences-first-manifest-as-problem-side-structural-differences}{%
\subsection{Exam Differences First Manifest as Problem-Side Structural Differences}\label{exam-differences-first-manifest-as-problem-side-structural-differences}}

A discussion of parallel-class fairness cannot start only from student mean scores or number of solved problems; it must first answer whether the exams have comparable structure on the problem side. The AI review shows that the 11 parallel-class exams differ in item count, high-difficulty ratio, and item-bank exposure. The overall difficulty of the 79 problems covers levels 1 to 5 (13/26/25/10/5 problems). By the mean of the raw AI overall difficulty, Class 15, Class 14, and Class 16 are in the higher-difficulty region; Class 10, Class 7, and Class 12 are in the lower-difficulty region. After introducing the item-bank exposure adjustment, the effective difficulty of Class 8 drops from 2.500 to 1.875, Class 9 from 2.615 to 2.462, and Class 2 from 2.333 to 2.167.

\textbf{Table 3 AI review exam difficulty and exposure adjustment}

\begin{longtable}[]{@{}
  >{\raggedright\arraybackslash}p{(\columnwidth - 12\tabcolsep) * \real{0.1111}}
  >{\raggedleft\arraybackslash}p{(\columnwidth - 12\tabcolsep) * \real{0.1481}}
  >{\raggedleft\arraybackslash}p{(\columnwidth - 12\tabcolsep) * \real{0.1481}}
  >{\raggedleft\arraybackslash}p{(\columnwidth - 12\tabcolsep) * \real{0.1481}}
  >{\raggedleft\arraybackslash}p{(\columnwidth - 12\tabcolsep) * \real{0.1481}}
  >{\raggedleft\arraybackslash}p{(\columnwidth - 12\tabcolsep) * \real{0.1481}}
  >{\raggedleft\arraybackslash}p{(\columnwidth - 12\tabcolsep) * \real{0.1481}}@{}}
\toprule\noalign{}
\begin{minipage}[b]{\linewidth}\raggedright
Exam
\end{minipage} & \begin{minipage}[b]{\linewidth}\raggedleft
Items
\end{minipage} & \begin{minipage}[b]{\linewidth}\raggedleft
Raw AI difficulty
\end{minipage} & \begin{minipage}[b]{\linewidth}\raggedleft
Adjusted AI difficulty
\end{minipage} & \begin{minipage}[b]{\linewidth}\raggedleft
High-difficulty count
\end{minipage} & \begin{minipage}[b]{\linewidth}\raggedleft
High-difficulty ratio
\end{minipage} & \begin{minipage}[b]{\linewidth}\raggedleft
Exposed items
\end{minipage} \\
\midrule\noalign{}
\endhead
\bottomrule\noalign{}
\endlastfoot
Class 15 & 8 & 3.375 & 3.375 & 4 & 0.500 & 0 \\
Class 14 & 7 & 3.143 & 3.143 & 3 & 0.429 & 0 \\
Class 16 & 7 & 3.143 & 3.143 & 2 & 0.286 & 0 \\
Class 9 & 13 & 2.615 & 2.462 & 2 & 0.154 & 3 \\
Class 13 & 5 & 2.400 & 2.400 & 1 & 0.200 & 0 \\
Class 4 & 8 & 2.375 & 2.375 & 1 & 0.125 & 0 \\
Class 12 & 6 & 2.333 & 2.333 & 0 & 0.000 & 0 \\
Class 2 & 6 & 2.333 & 2.167 & 1 & 0.167 & 1 \\
Class 7 & 6 & 2.167 & 2.167 & 0 & 0.000 & 0 \\
Class 8 & 8 & 2.500 & 1.875 & 1 & 0.125 & 8 \\
Class 10 & 5 & 1.600 & 1.600 & 0 & 0.000 & 0 \\
\end{longtable}

Table 3 shows that the AI review can decompose exam structure into several discussable dimensions. Class 9 does not have a high high-difficulty ratio, but its item count reaches 13, which may create significant time pressure; Class 15 and Class 14 have higher high-difficulty ratios, reflecting their ceiling-differentiation design; Class 8 illustrates that the intrinsic difficulty of problems and their effective difficulty in the exam context need to be explained separately.

\hypertarget{exam-level-relationships-are-directionally-consistent-but-can-only-serve-as-exploratory-evidence}{%
\subsection{Exam-Level Relationships Are Directionally Consistent but Can Only Serve as Exploratory Evidence}\label{exam-level-relationships-are-directionally-consistent-but-can-only-serve-as-exploratory-evidence}}

At the exam level, AI review difficulty and student whole-exam performance generally exhibit the expected direction: exams with higher AI difficulty mostly correspond to lower normalized student performance. For example, Class 16 has an adjusted AI difficulty of 3.143, a normalized student mean performance of 0.191, and a low-performance rate of 0.636; Class 10 has the lowest adjusted AI difficulty at 1.600, a normalized student mean performance of 0.716, and a full-completion rate of 0.336. There are also clear deviating cases: Class 15 has the highest adjusted AI difficulty (3.375), yet its normalized student mean performance reaches 0.477, which is related to that exam's structural design of ``ample support from basic problems, with high-difficulty problems responsible for ceiling differentiation'' (see Section 7.2), showing that the mean exam difficulty cannot be interpreted apart from the internal structure of the exam.

\textbf{Table 4 AI review adjusted difficulty versus student performance}

\begin{longtable}[]{@{}
  >{\raggedright\arraybackslash}p{(\columnwidth - 14\tabcolsep) * \real{0.1000}}
  >{\raggedleft\arraybackslash}p{(\columnwidth - 14\tabcolsep) * \real{0.1333}}
  >{\raggedright\arraybackslash}p{(\columnwidth - 14\tabcolsep) * \real{0.1000}}
  >{\raggedleft\arraybackslash}p{(\columnwidth - 14\tabcolsep) * \real{0.1333}}
  >{\raggedleft\arraybackslash}p{(\columnwidth - 14\tabcolsep) * \real{0.1333}}
  >{\raggedleft\arraybackslash}p{(\columnwidth - 14\tabcolsep) * \real{0.1333}}
  >{\raggedleft\arraybackslash}p{(\columnwidth - 14\tabcolsep) * \real{0.1333}}
  >{\raggedleft\arraybackslash}p{(\columnwidth - 14\tabcolsep) * \real{0.1333}}@{}}
\toprule\noalign{}
\begin{minipage}[b]{\linewidth}\raggedright
Exam
\end{minipage} & \begin{minipage}[b]{\linewidth}\raggedleft
Adjusted AI difficulty
\end{minipage} & \begin{minipage}[b]{\linewidth}\raggedright
Student metric
\end{minipage} & \begin{minipage}[b]{\linewidth}\raggedleft
Student mean
\end{minipage} & \begin{minipage}[b]{\linewidth}\raggedleft
Student median
\end{minipage} & \begin{minipage}[b]{\linewidth}\raggedleft
Normalized mean
\end{minipage} & \begin{minipage}[b]{\linewidth}\raggedleft
Full-completion/full-score rate
\end{minipage} & \begin{minipage}[b]{\linewidth}\raggedleft
Low-performance rate
\end{minipage} \\
\midrule\noalign{}
\endhead
\bottomrule\noalign{}
\endlastfoot
Class 15 & 3.375 & Solved count & 3.817 & 4.000 & 0.477 & 0.000 & 0.067 \\
Class 14 & 3.143 & Solved count & 2.253 & 2.000 & 0.322 & 0.000 & 0.189 \\
Class 16 & 3.143 & Solved count & 1.336 & 1.000 & 0.191 & 0.000 & 0.636 \\
Class 9 & 2.462 & Solved count & 3.344 & 2.000 & 0.257 & 0.000 & 0.312 \\
Class 13 & 2.400 & Score & 328.404 & 350.000 & 0.657 & 0.170 & 0.213 \\
Class 4 & 2.375 & Score & 431.974 & 440.000 & 0.540 & 0.026 & 0.243 \\
Class 12 & 2.333 & Solved count & 2.748 & 3.000 & 0.458 & 0.019 & 0.206 \\
Class 7 & 2.167 & Solved count & 2.857 & 3.000 & 0.476 & 0.000 & 0.214 \\
Class 2 & 2.167 & Score & 334.962 & 300.000 & 0.558 & 0.096 & 0.279 \\
Class 8 & 1.875 & Solved count & 3.557 & 4.000 & 0.445 & 0.016 & 0.148 \\
Class 10 & 1.600 & Solved count & 3.579 & 4.000 & 0.716 & 0.336 & 0.103 \\
\end{longtable}

To further observe ranking consistency, this paper computes exploratory correlations at the exam level.

\textbf{Table 5 Exploratory correlations between AI difficulty and student performance at the exam level}

\begin{longtable}[]{@{}
  >{\raggedright\arraybackslash}p{(\columnwidth - 10\tabcolsep) * \real{0.2105}}
  >{\raggedleft\arraybackslash}p{(\columnwidth - 10\tabcolsep) * \real{0.0702}}
  >{\raggedright\arraybackslash}p{(\columnwidth - 10\tabcolsep) * \real{0.2105}}
  >{\raggedright\arraybackslash}p{(\columnwidth - 10\tabcolsep) * \real{0.2456}}
  >{\raggedleft\arraybackslash}p{(\columnwidth - 10\tabcolsep) * \real{0.1228}}
  >{\raggedleft\arraybackslash}p{(\columnwidth - 10\tabcolsep) * \real{0.1404}}@{}}
\toprule\noalign{}
\begin{minipage}[b]{\linewidth}\raggedright
Sample
\end{minipage} & \begin{minipage}[b]{\linewidth}\raggedleft
n
\end{minipage} & \begin{minipage}[b]{\linewidth}\raggedright
AI difficulty index
\end{minipage} & \begin{minipage}[b]{\linewidth}\raggedright
Student metric
\end{minipage} & \begin{minipage}[b]{\linewidth}\raggedleft
Pearson r
\end{minipage} & \begin{minipage}[b]{\linewidth}\raggedleft
Spearman ρ
\end{minipage} \\
\midrule\noalign{}
\endhead
\bottomrule\noalign{}
\endlastfoot
All exams & 11 & Adjusted AI difficulty & Normalized student mean & -0.598 & -0.461 \\
All exams & 11 & Raw AI difficulty & Normalized student mean & -0.683 & -0.594 \\
Solved-count exams & 8 & Adjusted AI difficulty & Normalized student mean & -0.626 & -0.383 \\
Solved-count exams & 8 & Raw AI difficulty & Normalized student mean & -0.696 & -0.455 \\
\end{longtable}

Table 5 shows that the direction of the exam-level correlation generally matches the expectation that ``the higher the AI difficulty, the lower the student performance,'' but its strength is weaker than at the problem level (see Section 5.4) and can only be interpreted as exploratory evidence. There are three reasons. First, the sample size is only 11 exams. Second, student performance mixes two statistical bases, solved count and score, and exam item count, exam duration, class baseline, and item exposure may all affect student performance. Third, the model identity of the single reviewer carries a third-party endpoint boundary. Therefore, exam-level results are more suitable for raising risk signals to be reviewed, rather than serving directly as a basis for equating exam difficulty.

\hypertarget{the-exposure-discount-is-a-contextual-parameter-not-a-fitted-parameter}{%
\subsection{The Exposure Discount Is a Contextual Parameter, Not a Fitted Parameter}\label{the-exposure-discount-is-a-contextual-parameter-not-a-fitted-parameter}}

Item-bank exposure is an important contextual variable that distinguishes programming exams from ordinary paper-and-pencil tests. If students have already seen identical or highly similar problems before the exam, the effective exam difficulty of those problems decreases; but this decrease cannot be automatically identified by AI reading the problem statement. To test whether the 0.25 discount setting has a decisive effect on the conclusions, this paper further sets the exposure discount to 0, 0.10, 0.25, and 0.40, and recomputes the exam-level and problem-level Spearman correlations respectively.

\textbf{Table 6 Sensitivity analysis of the item-bank exposure discount}

\begin{longtable}[]{@{}
  >{\raggedleft\arraybackslash}p{(\columnwidth - 14\tabcolsep) * \real{0.1379}}
  >{\raggedright\arraybackslash}p{(\columnwidth - 14\tabcolsep) * \real{0.1034}}
  >{\raggedright\arraybackslash}p{(\columnwidth - 14\tabcolsep) * \real{0.1034}}
  >{\raggedleft\arraybackslash}p{(\columnwidth - 14\tabcolsep) * \real{0.1379}}
  >{\raggedright\arraybackslash}p{(\columnwidth - 14\tabcolsep) * \real{0.1034}}
  >{\raggedleft\arraybackslash}p{(\columnwidth - 14\tabcolsep) * \real{0.1379}}
  >{\raggedleft\arraybackslash}p{(\columnwidth - 14\tabcolsep) * \real{0.1379}}
  >{\raggedleft\arraybackslash}p{(\columnwidth - 14\tabcolsep) * \real{0.1379}}@{}}
\toprule\noalign{}
\begin{minipage}[b]{\linewidth}\raggedleft
Discount
\end{minipage} & \begin{minipage}[b]{\linewidth}\raggedright
Analysis level
\end{minipage} & \begin{minipage}[b]{\linewidth}\raggedright
Sample scope
\end{minipage} & \begin{minipage}[b]{\linewidth}\raggedleft
n
\end{minipage} & \begin{minipage}[b]{\linewidth}\raggedright
Predictor
\end{minipage} & \begin{minipage}[b]{\linewidth}\raggedleft
Outcome
\end{minipage} & \begin{minipage}[b]{\linewidth}\raggedleft
Pearson r
\end{minipage} & \begin{minipage}[b]{\linewidth}\raggedleft
Spearman ρ
\end{minipage} \\
\midrule\noalign{}
\endhead
\bottomrule\noalign{}
\endlastfoot
0.00 & Problem & All 79 problems & 79 & Adjusted AI difficulty & Pass rate & -0.831 & -0.871 \\
0.10 & Problem & All 79 problems & 79 & Adjusted AI difficulty & Pass rate & -0.822 & -0.859 \\
0.25 & Problem & All 79 problems & 79 & Adjusted AI difficulty & Pass rate & -0.796 & -0.857 \\
0.40 & Problem & All 79 problems & 79 & Adjusted AI difficulty & Pass rate & -0.756 & -0.780 \\
0.00 & Problem & Only 67 non-exposure-flagged problems & 67 & Adjusted AI difficulty & Pass rate & -0.825 & -0.866 \\
0.00 & Problem & All 79 problems & 79 & Adjusted AI difficulty & Non-attempt rate & 0.778 & 0.800 \\
0.10 & Problem & All 79 problems & 79 & Adjusted AI difficulty & Non-attempt rate & 0.761 & 0.765 \\
0.25 & Problem & All 79 problems & 79 & Adjusted AI difficulty & Non-attempt rate & 0.722 & 0.758 \\
0.40 & Problem & All 79 problems & 79 & Adjusted AI difficulty & Non-attempt rate & 0.671 & 0.650 \\
0.00 & Exam & All 11 exams & 11 & Mean adjusted AI difficulty & Normalized student mean & -0.683 & -0.594 \\
0.10 & Exam & All 11 exams & 11 & Mean adjusted AI difficulty & Normalized student mean & -0.657 & -0.465 \\
0.25 & Exam & All 11 exams & 11 & Mean adjusted AI difficulty & Normalized student mean & -0.598 & -0.461 \\
0.40 & Exam & All 11 exams & 11 & Mean adjusted AI difficulty & Normalized student mean & -0.527 & -0.282 \\
0.00 & Exam & Only 8 solved-count exams & 8 & Mean adjusted AI difficulty & Normalized student mean & -0.696 & -0.455 \\
0.10 & Exam & Only 8 solved-count exams & 8 & Mean adjusted AI difficulty & Normalized student mean & -0.676 & -0.431 \\
0.25 & Exam & Only 8 solved-count exams & 8 & Mean adjusted AI difficulty & Normalized student mean & -0.626 & -0.383 \\
0.40 & Exam & Only 8 solved-count exams & 8 & Mean adjusted AI difficulty & Normalized student mean & -0.564 & -0.287 \\
\end{longtable}

The sensitivity analysis yields results at three levels. First, the direction and strength of the problem-level correlation are generally stable: across the four discount levels, the Spearman ρ between adjusted AI difficulty and pass rate ranges from -0.780 to -0.871. Notably, the correlation is strongest when the discount is 0 (i.e., no adjustment), and it weakens monotonically as the discount increases; this is contrary to the assumption that ``the exposure discount improves explanatory power,'' suggesting that the raw difficulty of this AI review batch already corresponds well to student performance, and that the current discount setting may over-adjust on some exposed problems (see also point three below). The 67 non-exposure-flagged problems are unaffected by adjustment under any discount, and their correlation values are constant at Pearson r=-0.825 and Spearman ρ=-0.866, showing that the overall correlation over the 79 problems is not driven by the 12 exposed problems.

Second, the direction of the exam-level correlation is unchanged but its strength decreases as the discount increases: across all 11 exams, the Spearman ρ drops from -0.594 to -0.282. The correlation direction is negative under all four discount levels, consistent with theoretical expectation. The discount of 0.25 adopted in this paper is in the middle among the four discounts and is not the strongest-correlation setting, indicating that this choice is not a results-oriented post hoc optimization.

Third, the descriptive difference between exposed and non-exposure-flagged problems suggests that the exposure variable cannot be simply interpreted as unidirectionally lowering student difficulty. The mean raw AI difficulty of the 12 item-bank original problems is 2.667, slightly higher than the 2.582 of the 67 non-exposure-flagged problems; their student mean pass rate is 36.6\%, lower than the 42.8\% of the non-exposure-flagged problems, and their mean non-attempt rate is 49.6\%, higher than the 35.4\% of the non-exposure-flagged problems. That is, in this sample, exposed problems do not exhibit the overall pattern of ``seen before, hence easier,'' which is related to the fact that the exposed problems of Class 8 are concentrated in a single class and cannot be separated from the class effect. The exposure discount can therefore only serve as a contextualizing explanatory parameter, not a causal estimate of student performance.

In summary, the negative correlation between AI difficulty and student performance is directionally stable within the discount range of 0.00 to 0.40 and does not depend on the choice of a particular discount parameter; at the same time, under this review batch the exposure discount fails to increase the correlation strength, and its role should be positioned as ``explicitly recording the exposure context'' rather than ``improving prediction.''

\hypertarget{problem-level-behavioral-data-form-an-external-criterion}{%
\subsection{Problem-Level Behavioral Data Form an External Criterion}\label{problem-level-behavioral-data-form-an-external-criterion}}

At the exam level the analysis sample comprises only 11 exams, and it mixes score-based and solved-count-based grading. Stronger validity evidence comes from problem-level behavioral data: if the problems that AI review rates as harder not only have lower pass rates but also exhibit more non-attempts, delayed AC, or specific error types, then AI difficulty gains multiple external criteria at the problem level. Taking the 79 problems as the unit of analysis, this paper jointly analyzes AI review difficulty, item-bank exposure variables, and OpenJudge problem-level student behavioral statistics.

\textbf{Table 7. Problem-level correlations between AI review difficulty and student behavioral metrics for the 79 problems}

\begin{longtable}[]{@{}
  >{\raggedright\arraybackslash}p{(\columnwidth - 10\tabcolsep) * \real{0.1429}}
  >{\raggedleft\arraybackslash}p{(\columnwidth - 10\tabcolsep) * \real{0.1905}}
  >{\raggedright\arraybackslash}p{(\columnwidth - 10\tabcolsep) * \real{0.1429}}
  >{\raggedright\arraybackslash}p{(\columnwidth - 10\tabcolsep) * \real{0.1429}}
  >{\raggedleft\arraybackslash}p{(\columnwidth - 10\tabcolsep) * \real{0.1905}}
  >{\raggedleft\arraybackslash}p{(\columnwidth - 10\tabcolsep) * \real{0.1905}}@{}}
\toprule\noalign{}
\begin{minipage}[b]{\linewidth}\raggedright
Sample
\end{minipage} & \begin{minipage}[b]{\linewidth}\raggedleft
n
\end{minipage} & \begin{minipage}[b]{\linewidth}\raggedright
AI review metric
\end{minipage} & \begin{minipage}[b]{\linewidth}\raggedright
Student problem-level metric
\end{minipage} & \begin{minipage}[b]{\linewidth}\raggedleft
Pearson r
\end{minipage} & \begin{minipage}[b]{\linewidth}\raggedleft
Spearman ρ
\end{minipage} \\
\midrule\noalign{}
\endhead
\bottomrule\noalign{}
\endlastfoot
All problems & 79 & Raw AI overall difficulty & Pass rate & -0.831 & -0.871 \\
All problems & 79 & Exposure-adjusted AI difficulty & Pass rate & -0.796 & -0.857 \\
All problems & 79 & Raw AI overall difficulty & Student failure rate (1 − pass rate) & 0.831 & 0.871 \\
All problems & 79 & Exposure-adjusted AI difficulty & Non-attempt rate & 0.722 & 0.758 \\
Non-exposure-flagged problems & 67 & Raw AI overall difficulty & Pass rate & -0.825 & -0.866 \\
\end{longtable}

Table 7 shows a strong negative correlation between AI review difficulty and actual student problem pass rates. The Spearman correlation between raw AI overall difficulty and student pass rate is -0.871 (p\textless0.001, Fisher-approximate 95\% CI {[}-0.916, -0.805{]}), and -0.857 after exposure adjustment. After removing the 12 item-bank original problems, the Spearman correlation between raw AI difficulty and pass rate is still -0.866, indicating that the relationship is not driven by exposed item-bank problems. The problem-level correlation strength approaches the -0.905 of the first-stage answering-based experiment on 8 problems, showing that the review-based scale retains ranking consistency with student performance even after nearly tenfold expansion in problem coverage.

\begin{figure}
\centering
\includegraphics{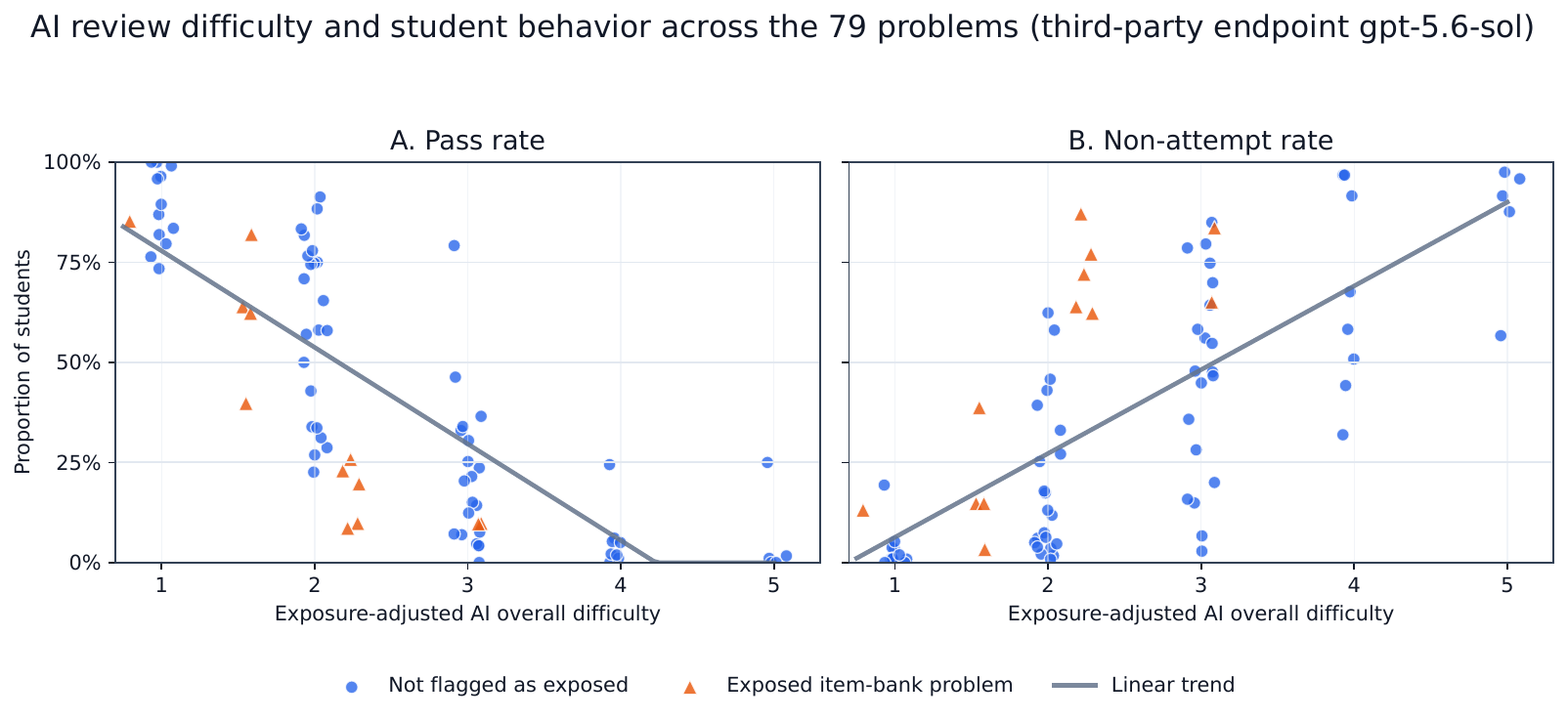}
\caption{Relationship between AI review difficulty and student pass rate and non-attempt rate for the 79 problems}
\end{figure}

Figure 1 visualizes the correlations in Table 7 as two oppositely directed trends: as exposure-adjusted AI difficulty increases, student pass rate declines overall, while non-attempt rate rises overall. The orange points in the figure denote exposed item-bank original problems; they do not simply fall in the ``high pass rate, low non-attempt rate'' region, suggesting that the exposure variable requires separate explanation.

The non-attempt rate provides process evidence beyond the pass rate. The Spearman correlation between raw AI difficulty and non-attempt rate is 0.800, indicating that for problems the AI judges to be harder, students are also more likely to leave the problem unattempted or give up. This result is especially important for programming exams: the difficulty of a problem is reflected not only in whether a submission achieves AC, but also in whether students are willing to invest an attempt within the limited exam time. Incorporating the non-attempt rate into the criterion advances the interpretation of difficulty from ``can it be solved'' to ``is it worth investing in solving within exam time.'' Note that item order and problem difficulty are positively correlated in most exams, and the non-attempt rate is affected by both item order and time budget, so it cannot be directly given a causal ``give-up mechanism'' interpretation.

The distribution of time to first AC provides evidence along the completion-speed dimension. The 79-problem sample contains 1,946 problem-level first-AC records, of which 50.3\% occur within 30 minutes after the exam starts and 72.9\% within 60 minutes. Grouped by AI overall difficulty, 83.7\% of first ACs for Level-1 problems occur within 30 minutes; Level-4 and Level-5 problems have no first AC within 30 minutes, and first ACs for Level-4 problems concentrate in the 90-to-120-minute interval (55.0\%). The AI difficulty grouping forms a clear gradient correspondence with student completion time.

A problem-level regression shows that, taking the student failure rate (1 − pass rate) as the dependent variable, each 1-level increase in raw AI overall difficulty raises the student failure rate by 0.251 on average, with a univariate model R²=0.690. After adding the exposure variable, the AI difficulty coefficient is 0.251 and the exposure variable coefficient is 0.040, with R²=0.692. In a test using CR1 robust standard errors clustered by exam, the AI difficulty coefficient remains significant (t=12.2); after adding exam fixed effects and item order, the coefficient is 0.232 and still significant, and leave-one-exam-out estimation yields coefficients ranging from 0.245 to 0.276. Because the distribution of exposed problems is imbalanced and observations are nested within 11 exams, these regressions remain correlational descriptions rather than causal estimates.

\textbf{Table 8. Student behavioral performance grouped by AI overall difficulty}

\begin{longtable}[]{@{}
  >{\raggedleft\arraybackslash}p{(\columnwidth - 8\tabcolsep) * \real{0.2000}}
  >{\raggedleft\arraybackslash}p{(\columnwidth - 8\tabcolsep) * \real{0.2000}}
  >{\raggedleft\arraybackslash}p{(\columnwidth - 8\tabcolsep) * \real{0.2000}}
  >{\raggedleft\arraybackslash}p{(\columnwidth - 8\tabcolsep) * \real{0.2000}}
  >{\raggedleft\arraybackslash}p{(\columnwidth - 8\tabcolsep) * \real{0.2000}}@{}}
\toprule\noalign{}
\begin{minipage}[b]{\linewidth}\raggedleft
AI overall difficulty
\end{minipage} & \begin{minipage}[b]{\linewidth}\raggedleft
Items
\end{minipage} & \begin{minipage}[b]{\linewidth}\raggedleft
Mean pass rate
\end{minipage} & \begin{minipage}[b]{\linewidth}\raggedleft
Mean non-attempt rate
\end{minipage} & \begin{minipage}[b]{\linewidth}\raggedleft
Erroneous-submission share
\end{minipage} \\
\midrule\noalign{}
\endhead
\bottomrule\noalign{}
\endlastfoot
1 & 13 & 88.3\% & 3.9\% & 64.6\% \\
2 & 26 & 59.6\% & 19.5\% & 71.8\% \\
3 & 25 & 20.5\% & 51.8\% & 83.7\% \\
4 & 10 & 6.5\% & 68.7\% & 94.8\% \\
5 & 5 & 5.5\% & 85.8\% & 87.6\% \\
\end{longtable}

Table 8 shows that as AI overall difficulty rises from Level 1 to Level 5, the mean student pass rate decreases monotonically (from 88.3\% to 5.5\%), the mean non-attempt rate increases monotonically (from 3.9\% to 85.8\%), and the erroneous-submission share rises overall (Level 5 is slightly lower than Level 4 because the total submission volume for Level-5 problems is very small). The AI review levels correspond to a continuous set of exam behavior patterns ranging from ``generally completed'' to ``generally abandoned.''

\begin{figure}
\centering
\includegraphics{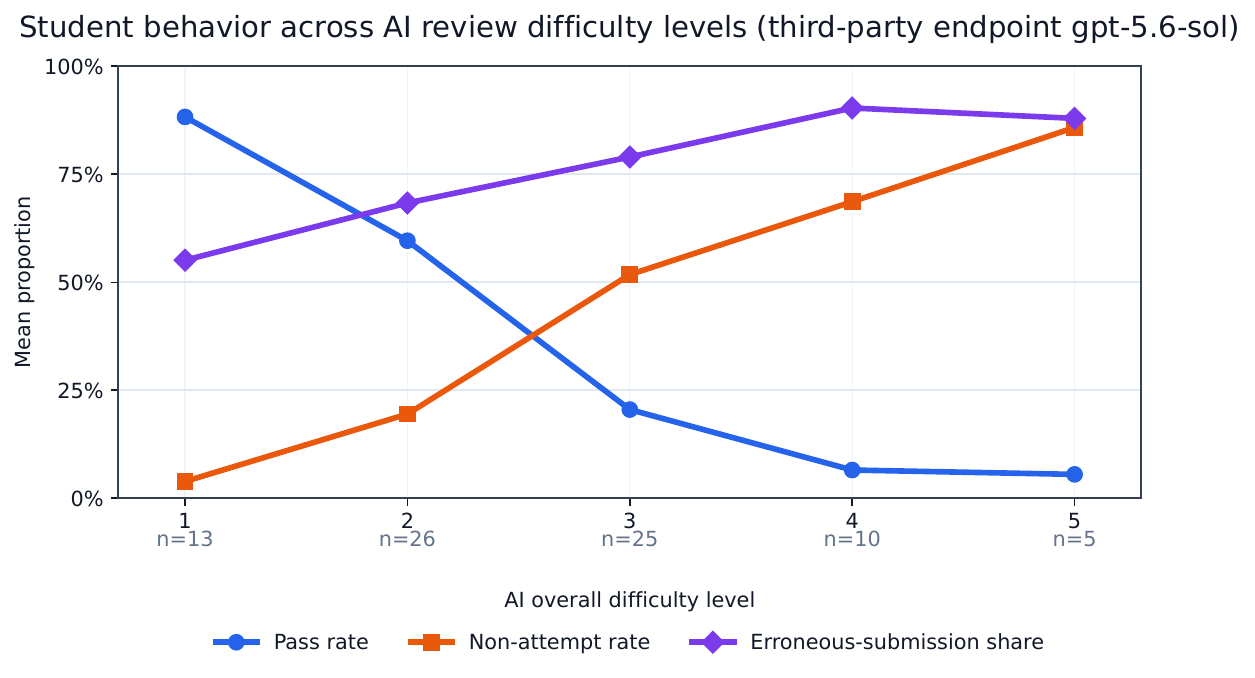}
\caption{Changes in student behavior across AI review difficulty levels}
\end{figure}

Figure 2 further groups the 79 problems by AI overall difficulty, showing the shape of how student behavior changes with difficulty level. The pass rate declines continuously from low to high levels, while the non-attempt rate and the erroneous-submission share rise overall; the non-attempt rate is notably higher for AI difficulty Levels 4 and 5 in particular.

\hypertarget{error-type-structure-supports-mechanism-level-mutual-explanation}{%
\subsection{Error-Type Structure Supports Mechanism-Level Mutual Explanation}\label{error-type-structure-supports-mechanism-level-mutual-explanation}}

If AI review is not only correlated with the pass rate but can also mutually explain the error-type structure, it can enter problem risk diagnosis. Problem-level status data show that, among the 79 problems, the dominant error type is Wrong Answer for 65 problems, Runtime Error for 8 problems, Time Limit Exceeded for 4 problems, and Compile Error for 2 problems. The dominance of Wrong Answer suggests that student failures may involve understanding the problem statement, algorithmic state design, boundary conditions, and implementation details; problems where Runtime Error and Time Limit Exceeded concentrate correspond respectively to implementation-risk and complexity-risk signals. At the dimension level, the Spearman correlation between the AI complexity-risk score and the student TLE-submission share is 0.350, the strongest relationship with TLE among the six review dimensions; this is directionally consistent with the design intent that ``the complexity-risk dimension captures the timeout risk of naive algorithms,'' but its strength is only moderate, indicating that the dimension score can serve only as a risk signal awaiting teacher review.

Observing specific cases, Class 14's ``Cow Election (data-augmented version),'' Class 7's ``Dynamic Graph Connectivity,'' and Class 15's ``Ask for Likes'' have the highest TLE-submission shares (0.43-0.57) and are best interpreted as complexity-risk-type problems; Class 16's ``Ice-Cream Reward Plan,'' Class 14's ``Chemical Formula,'' and Class 9's ``Weiming Lake Walk Check-in'' have relatively high Runtime Error proportions and are best interpreted as implementation-robustness-risk-type problems; Class 9's ``Currency Exchange,'' Class 16's ``Scrolls and Magic Stones,'' and Class 4's ``Maximize Remaining String'' have relatively high Wrong Answer proportions, indicating that these problems are more likely to cause errors in modeling, condition extraction, or detail judgment.

The error-type structure thereby forms an operational path for reviewing AI review dimensions: if the review assigns a high complexity risk while the student status shows a high TLE proportion, the teacher can further judge whether the two correspond to the same mechanism; the implementation or debugging dimensions can likewise be cross-checked against the RE, CE, and WA structure. This three-way cross-check of ``AI dimension score - student error structure - teacher mechanism judgment'' is the concrete working form in which the AI scale enters problem-setting review.

\hypertarget{longitudinal-review-based-calibration-results-transfer-and-boundaries-of-a-single-scale}{%
\section{Longitudinal Review-Based Calibration Results: Transfer and Boundaries of a Single Scale}\label{longitudinal-review-based-calibration-results-transfer-and-boundaries-of-a-single-scale}}

The cross-sectional parallel-class comparison mainly answers the question of comparability across the exam papers of different teachers in the same semester. Course quality improvement also requires answering another question: whether the difficulty structure of the same teacher's exam papers changes across different semesters and different course levels, and whether it can be tracked with the same scale. The longitudinal sample serves two functions in this paper: first, to test the feasibility of AI review as a cross-semester item-risk tracking tool, and second, to examine when the same procedure weakens across different course levels. To this end, this paper treats separately two sets of past final machine exams, from Data Structures and Algorithms B and from Introduction to Computing B. The former covers 4 final machine exams in Spring 2024, Spring 2025, Fall 2025, and Spring 2026, comprising 26 items in total; the latter covers 16 final machine exams from Fall 2008 to Fall 2025, comprising 106 items in total. Both sets of samples use exactly the same review protocol and batch as the cross-sectional sample. Table 9 gives the AI review difficulty and the mean item-level student performance for each exam of Data Structures and Algorithms B.

\textbf{Table 9 Longitudinal AI review difficulty and student performance across the same teacher's 4 final machine exams}

\begin{longtable}[]{@{}
  >{\raggedright\arraybackslash}p{(\columnwidth - 16\tabcolsep) * \real{0.0882}}
  >{\raggedleft\arraybackslash}p{(\columnwidth - 16\tabcolsep) * \real{0.1176}}
  >{\raggedleft\arraybackslash}p{(\columnwidth - 16\tabcolsep) * \real{0.1176}}
  >{\raggedleft\arraybackslash}p{(\columnwidth - 16\tabcolsep) * \real{0.1176}}
  >{\raggedleft\arraybackslash}p{(\columnwidth - 16\tabcolsep) * \real{0.1176}}
  >{\raggedleft\arraybackslash}p{(\columnwidth - 16\tabcolsep) * \real{0.1176}}
  >{\raggedleft\arraybackslash}p{(\columnwidth - 16\tabcolsep) * \real{0.1176}}
  >{\raggedleft\arraybackslash}p{(\columnwidth - 16\tabcolsep) * \real{0.1176}}
  >{\raggedright\arraybackslash}p{(\columnwidth - 16\tabcolsep) * \real{0.0882}}@{}}
\toprule\noalign{}
\begin{minipage}[b]{\linewidth}\raggedright
Semester
\end{minipage} & \begin{minipage}[b]{\linewidth}\raggedleft
Items
\end{minipage} & \begin{minipage}[b]{\linewidth}\raggedleft
Students
\end{minipage} & \begin{minipage}[b]{\linewidth}\raggedleft
Mean AI difficulty
\end{minipage} & \begin{minipage}[b]{\linewidth}\raggedleft
High-difficulty count
\end{minipage} & \begin{minipage}[b]{\linewidth}\raggedleft
High-difficulty ratio
\end{minipage} & \begin{minipage}[b]{\linewidth}\raggedleft
Mean pass rate
\end{minipage} & \begin{minipage}[b]{\linewidth}\raggedleft
Mean non-attempt rate
\end{minipage} & \begin{minipage}[b]{\linewidth}\raggedright
Dominant error type
\end{minipage} \\
\midrule\noalign{}
\endhead
\bottomrule\noalign{}
\endlastfoot
Spring 2024 & 6 & 137 & 2.000 & 0 & 0.000 & 0.656 & 0.230 & WA:6 \\
Spring 2025 & 6 & 111 & 2.333 & 1 & 0.167 & 0.635 & 0.204 & WA:6 \\
Fall 2025 & 6 & 80 & 2.667 & 1 & 0.167 & 0.408 & 0.402 & WA:5; TLE:1 \\
Spring 2026 & 8 & 120 & 3.375 & 4 & 0.500 & 0.473 & 0.404 & WA:7; TLE:1 \\
\end{longtable}

Table 9 shows that the mean AI difficulty rises semester by semester from 2.000 in Spring 2024 to 3.375 in Spring 2026, and the high-difficulty ratio rises from 0 to 0.500, presenting a clear upward trend in item difficulty; the mean student pass rate declines overall along with it (0.656 in Spring 2024, 0.408 in Fall 2025), and the non-attempt rate rises. Spring 2026 is a noteworthy deviation: its mean AI difficulty and high-difficulty ratio are the highest of the four exams, yet the mean student pass rate rebounds to 0.473, which is related to the structural design of this paper providing sufficient support from basic items (see Section 7.2). This indicates that the unit of interpretation for longitudinal tracking should be ``difficulty structure'' rather than merely the mean difficulty.

On the item-level data of 26 items, overall AI difficulty is strongly negatively correlated with student pass rate (Pearson r=-0.795, Spearman ρ=-0.829) and strongly positively correlated with the non-attempt rate (Pearson r=0.848, Spearman ρ=0.883), both consistent in magnitude with the cross-sectional 79-item sample. At the exam-paper level across the 4 exams, the Spearman correlation between mean AI difficulty and mean student pass rate is -0.800 (Pearson r=-0.726); however, there are only 4 observations at the exam-paper level, so this value serves only as a descriptive reference and should not support strong statistical inference.

In terms of error types, among the 26 items the dominant error type for 24 items is Wrong Answer, and for 2 items it is Time Limit Exceeded. In Fall 2025, ``Dynamic Graph Connectivity'' is TLE-dominated, with an AI complexity-risk score of 4, a student pass rate of 0.200, and a non-attempt rate of 0.488; in Spring 2026, ``LLM Recursive Card Generation'' has an overall AI difficulty of 5, a student pass rate of 0, and a non-attempt rate of 0.975. Such cases show that the longitudinal data chain can juxtapose AI-dimension scores with student behavior, providing teachers with tracking clues about ``which item types successively constitute complexity risk or abandonment risk''; the specific pedagogical conclusions still require the teacher to review them together with the item-setting intent.

The Introduction to Computing B sample is not a simple enlargement of the Data Structures and Algorithms B sample, but a stricter boundary test. This course is dominated by basic syntax, simulation, strings, sorting, and simple dynamic programming, the item difficulty range is narrower, and student performance is also more susceptible to training coverage, year-to-year variation, and cohort differences. Table 10 gives the exam-level summary of the past final machine exams of Introduction to Computing B. For this sample, all 106 items across 16 exams obtained parsable ranking and status student performance.

\textbf{Table 10 AI review difficulty and student performance of the same teacher's past final machine exams of Introduction to Computing B}

\begin{longtable}[]{@{}
  >{\raggedright\arraybackslash}p{(\columnwidth - 16\tabcolsep) * \real{0.0882}}
  >{\raggedleft\arraybackslash}p{(\columnwidth - 16\tabcolsep) * \real{0.1176}}
  >{\raggedleft\arraybackslash}p{(\columnwidth - 16\tabcolsep) * \real{0.1176}}
  >{\raggedleft\arraybackslash}p{(\columnwidth - 16\tabcolsep) * \real{0.1176}}
  >{\raggedleft\arraybackslash}p{(\columnwidth - 16\tabcolsep) * \real{0.1176}}
  >{\raggedleft\arraybackslash}p{(\columnwidth - 16\tabcolsep) * \real{0.1176}}
  >{\raggedleft\arraybackslash}p{(\columnwidth - 16\tabcolsep) * \real{0.1176}}
  >{\raggedleft\arraybackslash}p{(\columnwidth - 16\tabcolsep) * \real{0.1176}}
  >{\raggedright\arraybackslash}p{(\columnwidth - 16\tabcolsep) * \real{0.0882}}@{}}
\toprule\noalign{}
\begin{minipage}[b]{\linewidth}\raggedright
Semester
\end{minipage} & \begin{minipage}[b]{\linewidth}\raggedleft
Items
\end{minipage} & \begin{minipage}[b]{\linewidth}\raggedleft
Students
\end{minipage} & \begin{minipage}[b]{\linewidth}\raggedleft
Mean AI difficulty
\end{minipage} & \begin{minipage}[b]{\linewidth}\raggedleft
High-difficulty count
\end{minipage} & \begin{minipage}[b]{\linewidth}\raggedleft
High-difficulty ratio
\end{minipage} & \begin{minipage}[b]{\linewidth}\raggedleft
Mean pass rate
\end{minipage} & \begin{minipage}[b]{\linewidth}\raggedleft
Mean non-attempt rate
\end{minipage} & \begin{minipage}[b]{\linewidth}\raggedright
Dominant error type
\end{minipage} \\
\midrule\noalign{}
\endhead
\bottomrule\noalign{}
\endlastfoot
Fall 2008 & 10 & 100 & 2.900 & 3 & 0.300 & 0.425 & 0.492 & WA:10 \\
Fall 2013 & 8 & 93 & 2.625 & 2 & 0.250 & 0.324 & 0.503 & WA:6; CE:2 \\
Fall 2014 & 8 & 180 & 2.375 & 1 & 0.125 & 0.449 & 0.391 & WA:8 \\
Fall 2016 & 6 & 150 & 2.667 & 1 & 0.167 & 0.393 & 0.361 & WA:6 \\
Fall 2017 & 6 & 155 & 3.333 & 2 & 0.333 & 0.326 & 0.446 & WA:6 \\
Fall 2018 & 6 & 141 & 2.833 & 2 & 0.333 & 0.486 & 0.342 & WA:4; RE:1; TLE:1 \\
Fall 2019 & 6 & 136 & 2.667 & 1 & 0.167 & 0.533 & 0.332 & WA:4; RE:2 \\
Fall 2020 (domestic) & 6 & 88 & 2.333 & 1 & 0.167 & 0.674 & 0.229 & WA:6 \\
Fall 2020 (international) & 7 & 29 & 2.571 & 2 & 0.286 & 0.345 & 0.527 & WA:6; CE:1 \\
Fall 2021 (domestic) & 6 & 156 & 2.667 & 1 & 0.167 & 0.690 & 0.194 & WA:5; RE:1 \\
Fall 2021 (international) & 6 & 34 & 2.500 & 1 & 0.167 & 0.377 & 0.475 & WA:5; RE:1 \\
Fall 2022 (domestic) & 6 & 188 & 2.833 & 2 & 0.333 & 0.647 & 0.253 & WA:6 \\
Fall 2022 (international) & 6 & 21 & 2.833 & 2 & 0.333 & 0.302 & 0.627 & WA:4; CE:1 \\
Fall 2023 & 7 & 184 & 2.857 & 2 & 0.286 & 0.516 & 0.355 & WA:4; MLE:1; PE:1; TLE:1 \\
Fall 2024 & 6 & 178 & 2.833 & 2 & 0.333 & 0.654 & 0.209 & WA:5; RE:1 \\
Fall 2025 & 6 & 189 & 2.833 & 1 & 0.167 & 0.616 & 0.238 & WA:6 \\
\end{longtable}

Introduction to Computing B presents an evidence landscape markedly different from that of Data Structures and Algorithms B. First, AI review does not exhibit scale compression in this course: the overall difficulty of the 106 items covers levels 1 to 5 (8/41/31/24/2 items), and 26 items are rated as high-difficulty. Second, at the item level the Pearson/Spearman between overall AI difficulty and student pass rate is -0.550/-0.552, and with the non-attempt rate is 0.541/0.564---the correlation direction is correct and the strength is moderate, clearly weaker than the -0.86/-0.83 magnitude of the two Data Structures and Algorithms B samples. Finally, at the exam-paper level across the 16 exams, the Spearman correlation between mean AI difficulty and mean student pass rate is only -0.063 (Pearson r=-0.168), close to zero. That is, the AI scale can still form a moderate-strength ordering signal for single-item difficulty in an introductory course, but can no longer explain overall performance differences at the exam-paper level; the contrast between the ``domestic/international'' papers of the same year in Table 10 directly explains the reason---the two papers have highly overlapping items and close mean AI difficulty, yet the mean student pass rate differs by about a factor of two (e.g., 0.690 versus 0.377 in Fall 2021), and cohort differences rather than item difficulty dominate exam-paper-level performance.

Because the international-student papers extensively reuse items from the domestic-student papers, among the 106 input items there are 13 duplicate pairs with identical item statements and reference code, and in 5 of these groups this review batch gave overall difficulties differing by 1 level. Recomputing under three treatments---replacing with the mean of the duplicate group, within-group swap, and retaining only one item per group (n=93)---the Spearman between AI difficulty and pass rate is -0.552, -0.538, and -0.535 respectively, essentially consistent with the full sample's -0.552, indicating that jitter in the review output does not alter the above conclusion; at the same time, this again reminds us that \texttt{temperature=0} does not amount to output determinism.

The contrast between the two courses shows that the explanatory power of the AI review scale is related to the course level and the level of analysis: it is strongest at the item level in algorithm-oriented courses, weakens at the item level in the introductory course, and fails at the exam-paper level in the introductory course. The main source of the boundary is not the AI scoring scale, but the dominant role that non-item factors---such as student cohort composition and the historical training environment in the introductory course---play in performance. Absolute score values across different courses also cannot be compared directly without anchoring.

\begin{figure}
\centering
\includegraphics{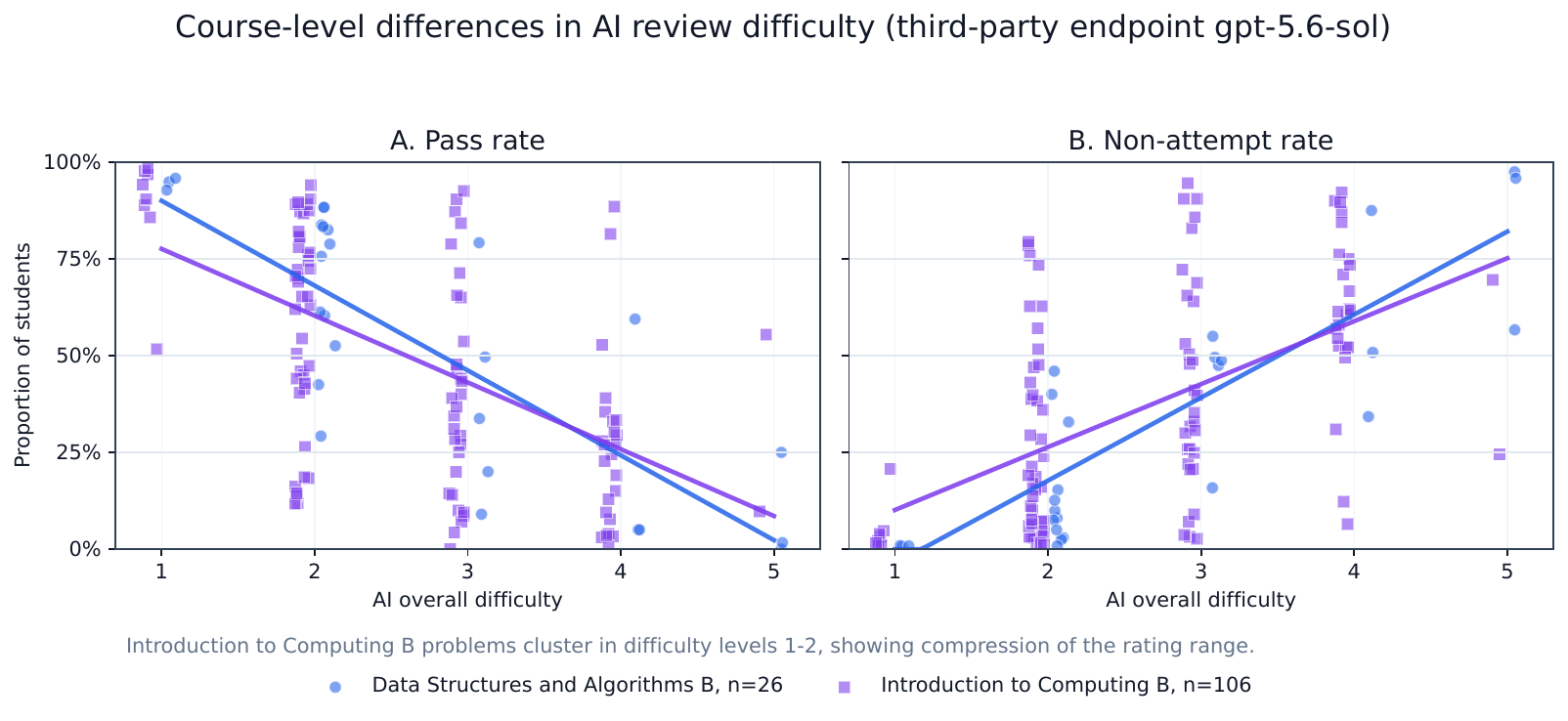}
\caption{AI review difficulty distribution and student performance in Data Structures and Algorithms B and Introduction to Computing B}
\end{figure}

Figure 3 places the longitudinal items of the two courses in the same AI review coordinate system. The AI score distributions of both courses cover multiple difficulty levels, but in Introduction to Computing B the dispersion of student performance corresponding to the same AI difficulty level is markedly larger, intuitively presenting the above boundary form of ``item-level signal preserved, exam-paper-level explanation failing.''

\hypertarget{illustrative-cases-and-educational-interpretation}{%
\section{Illustrative Cases and Educational Interpretation}\label{illustrative-cases-and-educational-interpretation}}

\hypertarget{high-ai-difficulty-and-low-student-performance-class-16}{%
\subsection{High AI Difficulty and Low Student Performance: Class 16}\label{high-ai-difficulty-and-low-student-performance-class-16}}

Class 16's adjusted AI difficulty is 3.143, tied for the second highest among the 11 papers, and its normalized mean student performance is 0.191 with a low-performance rate of 0.636, both the weakest among the 11 papers. This paper has 7 items, and AI review shows that both its conceptual difficulty and implementation difficulty are relatively high (the mean implementation difficulty of 3.429 is the highest among the classes), including 2 high-difficulty items. This case shows that paper difficulty comes not only from the item count but also from the combined load of the items in modeling, implementation, and reading comprehension. When low-to-medium-difficulty items provide insufficient support, high-difficulty items may cause a relatively high proportion of students to remain in the low-pass-count range.

\hypertarget{high-difficulty-ratio-and-discriminating-top-students-class-15}{%
\subsection{High-Difficulty Ratio and Discriminating Top Students: Class 15}\label{high-difficulty-ratio-and-discriminating-top-students-class-15}}

Class 15 contains 8 items, of which 4 are high-difficulty, giving a high-difficulty ratio of 0.500 and an adjusted AI difficulty of 3.375, both the highest among the 11 papers. On the student side, among the 120 actual participants, the mean number of items passed is 3.817, the median is 4, and the low-performance rate is only 0.067. The paper with the highest AI difficulty simultaneously maintains a relatively high student completion level, indicating that its basic and medium items can support most students in achieving a certain level of completion, while high-difficulty items are used to discriminate top students. This case reflects a relatively clear structural division of labor between ``basic coverage'' and ``discriminating top students,'' and is also the main source of the deviation in Section 5.2, where the exam-paper-level correlation is weaker than the item-level correlation: two papers with the same mean difficulty can produce completely different distributions of student performance because of different structural divisions of labor.

\hypertarget{item-count-pressure-class-9}{%
\subsection{Item-Count Pressure: Class 9}\label{item-count-pressure-class-9}}

The Class 9 paper contains 13 items, the largest item count among the 11 papers. Of its 13 items, 3 are original items from the item bank; the mean raw AI difficulty is 2.615, and after exposure adjustment it is 2.462, with only 2 high-difficulty items (a ratio of 0.154), yet the normalized mean student performance is only 0.257 and the median number of items passed is 2. This case suggests that paper difficulty cannot be fully explained by the mean single-item difficulty. The item count, the switching cost between items, the reading burden, and the pressure of time allocation themselves constitute sources of difficulty at the exam-paper level.

\hypertarget{item-bank-exposure-12-original-items-across-three-papers}{%
\subsection{Item-Bank Exposure: 12 Original Items Across Three Papers}\label{item-bank-exposure-12-original-items-across-three-papers}}

This study annotates a total of 12 original items from the item bank, distributed across Class 8, Class 9, and Class 2. All 8 items of Class 8 are original items from the practice item bank, with a raw AI difficulty of 2.500, which drops to 1.875 after exposure adjustment; the 3 original items of Class 9 reduce the whole paper's adjusted difficulty from 2.615 to 2.462; the 1 original item of Class 2 reduces the whole paper's adjusted difficulty from 2.333 to 2.167. This case shows that whether an item has been seen before and whether it comes from an in-course practice system will change the effective difficulty and the evaluative meaning of the paper. Combined with the sensitivity analysis in Section 5.3, the exposure discount of 0.25 is a conservative setting, and within the range of 0 to 0.40 it does not change the correlation direction; however, under this review batch the discount does not improve the item-level correlation, suggesting that the value of the discount lies in explicitly recording the exposure context rather than in improving difficulty prediction. For the course team, original items from the item bank are not necessarily unusable in exams, but the exposure type and the basis for the discount should be explicitly recorded in score interpretation and paper comparison.

\hypertarget{discussion}{%
\section{Discussion}\label{discussion}}

Around the four research questions, the results of this paper can be summarized in four points. First, AI solving-based calibration in a real online-judge environment can produce a problem ordering that is highly consistent with student pass rates (RQ1), showing that the solving process of large language models can serve as external evidence for explaining the difficulty of programming exams. Second, auditable API review, at the problem level of the 79-problem cross-sectional sample, forms strong correlations with student pass rate and non-attempt rate (Spearman ρ=-0.871 and 0.800), and is mutually corroborated with first-AC time and error-type structure; at the exam level the direction is consistent but the strength is weaker, so it can only serve as exploratory evidence (RQ2). Third, item-bank exposure, item-quantity pressure, non-attempt rate, and error type effectively explain the deviant cases and the applicability boundary of the AI ruler: under the present review batch the exposure discount did not improve the correlation, item quantity and the division of labor in exam structure are the main sources of exam-level deviation, and in Introduction to Computer Science B the near-zero exam-level correlation constitutes a course-level boundary (RQ3). Fourth, under the same review protocol, the AI ruler supports longitudinal tracking of the difficulty structure of the same teacher's Data Structures and Algorithms B exams (26-problem problem-level Spearman ρ=-0.829), while in Introduction to Computer Science B it degrades to a medium problem-level signal (ρ=-0.552) and fails at the exam level (RQ4). In whichever scenario, the AI difficulty ruler is only suitable for entering the evidence discussion of the teachers' item-setting community, and is not suitable for being converted into automatic score adjustment or individual student evaluation rules; its evidentiary status is further constrained by the boundary of third-party endpoint model identity and by the single-reviewer design.

\hypertarget{from-model-capability-evaluation-to-course-exam-evaluation}{%
\subsection{From Model Capability Evaluation to Course Exam Evaluation}\label{from-model-capability-evaluation-to-course-exam-evaluation}}

One core shift in this paper is turning large language models from the object of evaluation into an auxiliary tool for evaluation. Benchmarks such as HumanEval, MBPP, and CodeContests are mainly used to measure the code-generation capability of models, whereas real programming courses care more about whether a problem fits the teaching objectives, whether an exam has a reasonable gradient, and whether the scores of different parallel classes have a basis for interpretation. The first-stage answering experiment of this paper shows that the solving process of large language models can provide a reviewable external reference for these questions; the three groups of auditable API review further show that reading-based review can likewise provide, at the problem level, a reference highly consistent with student performance, with its boundary mainly appearing at the exam level in introductory courses.

\hypertarget{complementarity-of-solving-based-calibration-and-review-based-calibration}{%
\subsection{Complementarity of Solving-Based Calibration and Review-Based Calibration}\label{complementarity-of-solving-based-calibration-and-review-based-calibration}}

Solving-based calibration is closest to the real problem-solving process; it can produce fine-grained evidence such as pass rate, number of attempts, running time, and failure feedback, and is suitable for key exams, representative problems, and item-setting review. Review-based calibration does not require the AI to actually submit code, can rapidly cover more exams, and is suitable for cross-sectional comparison of parallel classes, pre-item-setting screening, and benchmark construction. A more robust path is to combine the two: use multi-model answering experiments to establish the credibility of the ruler, use the review-based process to expand coverage, and then calibrate with student-group performance and teacher judgment.

The problem-level results in Sections 5.4 and 5.5, show that AI review difficulty is not only strongly correlated with the pass rate of the 79 problems, but also forms directionally consistent multiple correspondences with the non-attempt rate, the distribution of first-AC time, and the error-type structure; among them the correspondence between the complexity-risk dimension and the proportion of TLE submissions (Spearman ρ=0.350) preliminarily shows that dimension scoring has explanatory potential at the mechanistic level, but the strength is only medium, and whether each dimension stably corresponds to different student behavior signals still needs to be tested with multiple reviewers and more course data. The longitudinal results in Section 6 show that the same review protocol can run across semesters and across courses while retaining problem-level explanatory power, providing feasibility evidence for tracking item-setting quality.

The teacher's experiential review plays an irreplaceable explanatory role in this process. API review can output structured dimensions, but the repeated inputs in this batch have already shown that it is not fully stable; online-judge data can provide student-group performance, yet ``why does a problem produce such performance in this class'' still needs to be understood by returning to the course objectives, training coverage, classroom emphasis, problem source, and item-setting intent. In this paper, the first author, as the instructor of Class 15, has direct item-setting and post-exam review experience for this class's 8-problem exam and for similar exams over the years, but this interpretation is still limited by a single perspective. Follow-up research should introduce the instructors of each parallel class for independent review, forming multi-teacher explanatory validation in the true sense.

\hypertarget{the-evidence-chain-for-fairness-in-parallel-class-evaluation}{%
\subsection{The Evidence Chain for Fairness in Parallel-Class Evaluation}\label{the-evidence-chain-for-fairness-in-parallel-class-evaluation}}

Parallel-class exam comparison cannot rely solely on average score or number of problems passed. Student scores are simultaneously affected by exam difficulty, class foundation, teaching process, problem exposure, exam time, and item-quantity structure. The value of the AI difficulty ruler lies in adding a group of problem-side and exam-side evidence, enabling the course group to discuss ``why does this exam perform relatively low,'' ``is the problem itself harder, or is the difference caused by item quantity and exposure factors,'' and ``which problems are used for basic evaluation and which are used to distinguish high-level students.'' Longitudinal comparison can in turn help teachers track ``whether this semester emphasizes hard problems more than in the past,'' ``whether the support of basic problems is insufficient,'' and ``which problem types continuously cause high non-attempt rate or TLE risk.'' Such an evidence chain combining the cross-sectional and the longitudinal is more in line with the professional requirements of educational evaluation for test fairness than ranking based purely on scores {[}35{]}.

However, fairness judgment cannot stop at the technical level of ``exam comparability.'' The question that truly needs to be answered in teaching management is: within what range is the difficulty difference calibrated across exams acceptable? This study cannot give an acceptable threshold---the 11 exams have no common anchor items, common ability baseline, or equating design, and the review comes from a single reviewer whose model identity is bounded, so this paper can only conduct comparison of exam structure risk and cannot claim to have completed a parallel-class fairness test. What can be stated is only the analytic framework: after accumulating equated data across more courses and more semesters, the course group may then, based on the distribution of differences in mean review difficulty, delineate an empirical boundary between ``normal item-setting variation'' and ``structural bias that requires systematic examination of item quantity, the proportion of hard problems, and exposure variables.''

Parallel-class comparison also needs to be alert to a problem masked by the current aggregate data: the same exam may present a different difficulty structure to different student subgroups (such as those differing in programming foundation, OJ usage experience, or course-selection background). The 79-problem analysis of this paper is based on the whole-class pass rate and cannot distinguish subgroup differences---this is why the AI ruler, when currently discussing ``between-class fairness,'' has been unable to touch on ``within-class fairness.'' Follow-up research should collect finer-grained student-stratification data along this direction.

Item-bank exposure should be regarded as a structural variable in programming exam evaluation. The sensitivity analysis in Section 5.3, shows that the correlation direction is stable across the four discount levels, with the exposure discount of 0.25 lying in the middle position rather than being an extreme choice; at the same time, under the present review batch the discount did not improve the correlation strength, and the value of exposure records lies in contextual explanation rather than difficulty prediction. Therefore, when setting items, the course team should annotate each problem with the exposure type (fully original problem, adapted problem, new problem), item-bank visibility, and training coverage, and accordingly conduct contextualized interpretation in cross-sectional exam comparison. The specific value of the exposure discount can vary with the course and the item-bank scale, but the inclusion of the exposure concept itself should become a standard step of item-setting review.

\hypertarget{implications-for-the-assessment-objectives-of-programming-courses}{%
\subsection{Implications for the Assessment Objectives of Programming Courses}\label{implications-for-the-assessment-objectives-of-programming-courses}}

If mainstream large language models can stably solve a large number of routine programming problems, course assessment needs to rethink its capability objectives {[}33{]}{[}36{]}. Basic syntax, template reproduction, and routine data-structure implementation remain important, but exams should not remain for long at the level of mechanical coding that models can quickly complete. Future programming evaluation should place more emphasis on problem modeling, algorithm selection, complex-constraint analysis, boundary-data construction, code explanation, debugging strategy, and human-AI collaborative programming ability {[}34{]}{[}37{]}. The AI ruler can not only help teachers judge problem difficulty, but also inversely suggest the direction for adjusting course evaluation objectives.

\hypertarget{ethical-boundaries-and-governance-requirements}{%
\subsection{Ethical Boundaries and Governance Requirements}\label{ethical-boundaries-and-governance-requirements}}

The AI difficulty ruler must observe clear boundaries. These boundaries are not principle statements unrelated to the empirical findings, but operational requirements derived from concrete research results.

First, AI results can only be used for auxiliary analysis at the problem and exam levels, and cannot be directly used to evaluate individual student ability. The Introduction to Computer Science B sample has already suggested a concrete risk: in introductory courses and data with a long historical span, the exam-level correlation is near zero, and the same AI difficulty level corresponds to highly dispersed student performance. If used for individual evaluation in such scenarios, the error would far exceed the acceptable range.

Second, student data should be limited to anonymized group statistics, and names, student IDs, accounts, IPs, personal submission code, or personally identifiable information should not be released. Considering that during the AI review process the reference answers and problem-statement data may be transmitted via the API, before use one needs to confirm whether the data-processing agreement prohibits the platform from using the submitted data for model training. This shows that operational-level governance requirements also need supporting technical-level constraints.

Third, before problem statements, test data, and reference answers are made public, the scope of authorization needs to be confirmed. Fourth, the model version, prompt version, review date, endpoint, temperature, prompt hash, and scoring schema must be recorded together with the results. The concrete meaning of this requirement in this study is: the three groups of review batches used in this paper all come with complete source metadata and have passed provenance verification, but it must still be disclosed that the third-party endpoint cannot authenticate the identity of the upstream model; the baseline files generated early on by deterministic rules, lacking model-source metadata, can only be used for pipeline testing and are thus excluded from the evidence. If used later for longitudinal quality tracking, a monitoring and calibration mechanism for reviewer version drift needs to be established.

Fifth, the AI ruler should not automatically trigger score adjustment, and the final interpretation still needs to be completed jointly by teachers and the course group based on teaching objectives. But prohibitions alone are not enough---procedural safeguards need to be established at the same time: who may view AI review results, in what form they are presented, whether they are included in the agenda of formal teaching meetings, and whether item-setting teachers have a mechanism for objection. The verification process of this study once exposed a more fundamental misuse risk: if source metadata and genuine model records are missing, the output of a deterministic-rule script can easily be mis-recorded as AI review results---this is also the direct reason this paper insists on full-chain provenance verification. Therefore, clear review checkpoints must be set: the output of the AI ruler should always enter the course-group discussion in the form of risk markers (such as ``high difficulty,'' ``high TLE risk,'' ``high non-attempt-rate warning'') rather than precise scores, and the teacher's final judgment shall prevail.

Sixth, students should enjoy the right to be informed regarding AI-assisted evaluation. Without this link, ``human-AI collaboration'' is only a one-way teacher-side tool and does not constitute a complete governance loop.

\hypertarget{limitations-and-future-research}{%
\section{Limitations and Future Research}\label{limitations-and-future-research}}

This paper still has several limitations. First, the solving-based experiment has only 8 problems and 10 AI models; although the ranking correlation is significant, the number of problems is small, and the first stage still needs supplementary archiving of complete model versions, run protocols, OJ receipts, and raw answering records. Second, the review-based evidence comes from a single batch run of a single reviewer and is completed via a third-party OpenAI-compatible endpoint: the model identifier in the requests and returns (gpt-5.6-sol) cannot authenticate the upstream as an official OpenAI model, and the reviewer identity constitutes the most important source boundary of this paper. The content audit further shows that the projected student pass-rate interval in the review output corresponds mechanically to the overall difficulty and cannot serve as independent evidence; a few complexity texts are filled in according to the ideal solution rather than the reference code, and the cost calibration for large-integer arithmetic is inconsistent---although these free-text field problems do not affect the scoring matrix, they suggest that the textual explanation of the review output must be manually verified before use. The three-column archival comparison of the 12 problems (this API batch, the Opus 4.8 labeled archive, and the DeepSeek labeled archive) shows consistent ranking of the historical scoring matrices, but the old archives lack raw responses, run metadata, and protocol binding, and cannot serve as formal reliability evidence.

Third, the internal expert knowledge of the first author as the instructor of Class 15 helps explain the course context, but may also bring an item-setter's perspective bias; follow-up work should introduce several non-item-setting teachers or teaching assistants for cross-review. Fourth, the analysis of problem-level student behavior data involves a nested structure---79 problems nested within 11 exams. Cluster-robust CR1 standard errors clustered by exam and exam fixed-effect tests show that the AI difficulty coefficient remains significant across settings (0.232-0.251), and the coefficient range when removing single exams one by one is 0.245-0.276; but with only 11 clusters being too few, and with problem order and problem difficulty being confounded, the above tests can only serve as sensitivity evidence rather than strict inference.

Fifth, although the student behavior data already cover pass rate, non-attempt rate, number of submissions, distribution of first-AC time, and error type, they still lack finer process indicators such as student ability stratification, code similarity, and complete submission sequences (e.g., error transition paths), so a complete cognitive diagnosis or IRT model cannot yet be established. More importantly, the current aggregate analysis cannot answer the question ``whether the same exam presents the same difficulty structure to different student subgroups''---this is the key evidence needed for the AI ruler to move from between-class fairness to within-class fairness discussion. Sixth, the whole-exam student performance of the 11 exams mixes the two statistical calibers of number of problems passed and score, so the exam-level correlation analysis can still only serve as exploratory evidence; although the longitudinal sample expands to two courses, it still comes from the same teacher and is suitable for descriptive tracking and interpretive comparison, not for strong statistical inference.

Seventh, although the item-bank exposure discount has undergone sensitivity analysis (0.00--0.40), it is currently still a conservative empirical setting. The exposed problems are concentrated in Class 8 (fully exposed), and within that class the ``exposure effect'' and the ``class effect'' cannot be distinguished; more course and more exposure-type data are needed to separate the two. Eighth, the problem-level student behavior data of this paper depend on scraping the OpenJudge ranking/status pages; in some classes there is a slight difference between the statistics visible on the ranking page and the aggregated result buckets of the source documents, and this paper retains these differences in the data tables without forcibly making manual corrections.

Ninth, the potential training-data contamination problem. ChatGPT may have encountered the public OpenJudge problems and their solutions in its pretraining corpus. If its problem-solving performance is partly based on memorization rather than reasoning, then the logical basis of the first stage for selecting the reviewer as ``the only AK model and the only one that passed I30547'' would be weakened. The following factors mitigate the impact of the contamination hypothesis to some extent: the 10 models in the first stage answered in the same environment, and if ChatGPT's memory advantage were caused by contamination, it would be hard to explain why the other models did not reach the same level on all problems; review-based calibration outputs dimension scores based on the problem statement and reference answer, is independent of the solving process, and the path by which memorization affects the review is more indirect. It should be added that the reviewer of this batch is accessed via a third-party endpoint, and its degree of training-data overlap is harder to assess than through official channels. Despite the above mitigating factors, follow-up research should still, on an informed basis, assess the degree of training-data overlap and design de-contamination validation experiments.

Follow-up research can advance in four directions. First, continue to expand problem-level student process data, especially complete submission sequences, error transition paths, and student-ability-stratified performance. Second, introduce multiple AI reviewers and teacher-expert scoring, compute review consistency, and analyze which problem types are more prone to AI review disagreement. Third, expand to multi-semester, multi-course, multi-teacher samples, and adopt hierarchical models or IRT models to simultaneously estimate problem difficulty, class differences, teacher item-setting style, and model capability. Fourth, build an AI-calibration benchmark for educational-scenario programming exams, incorporating on-campus parallel-class exam problems, longitudinal exam problems over the years, problem-exposure annotations, AI answering-process data, AI review data, student-group statistics, online-judge process data, and external code-benchmark anchors into the same reviewable framework, so that the AI ruler gradually moves from single-course analysis toward reusable, comparable, and auditable educational-evaluation infrastructure.

\hypertarget{conclusion}{%
\section{Conclusion}\label{conclusion}}

Focusing on the problem of difficulty calibration for parallel-class programming exams, this paper proposes and tests a multi-evidence path for AI-assisted evaluation. Real synchronous machine exams show that the answering performance of 10 large language models can produce a problem-difficulty ordering highly consistent with student pass rates: AI pass rate and student pass rate have Spearman ρ=0.866, and the composite difficulty index and student pass rate have Spearman ρ=-0.905. On this basis, source-auditable single-reviewer structured review expanded problem coverage by an order of magnitude and maintained strong consistency with student performance at the problem level: in the 79-problem cross-sectional sample, AI overall difficulty and student pass rate have Spearman ρ=-0.871, and with non-attempt rate ρ=0.800, while in the 26-problem longitudinal sample the corresponding values are -0.829 and 0.883; the AI difficulty groups simultaneously correspond to continuous behavioral gradients in pass rate, non-attempt rate, first-AC time, and error type. The Introduction to Computer Science B sample, in contrast, delineates the boundary of the ruler: the problem-level correlation drops to medium (ρ=-0.552), the exam-level correlation across the 16 exams is near zero, and in introductory courses non-problem factors such as student-group composition dominate exam-level performance.

The validity of the above evidence is bounded by three source limits: the review is completed via a third-party OpenAI-compatible endpoint, and the model label cannot authenticate the upstream as an official OpenAI model; the reviewer is a single reviewer, and there is as yet no source-complete multi-reviewer reliability; the repeated-problem test shows that \texttt{temperature=0} does not constitute a guarantee of output determinism, but the scoring jitter does not change the main correlations. This paper therefore positions review-based AI evidence as a ``source-auditable, identity-bounded'' exploratory ruler, whose methodological premises---auditable API calls, complete source metadata, content verification, and reproducible validation---are themselves part of the research conclusion.

The research conclusion also delimits the boundary of use for the AI ruler. The AI difficulty ruler is only suitable for serving item-setting review, exam-structure analysis, problem-level risk diagnosis, parallel-class fairness discussion, longitudinal quality tracking, and benchmark construction; it cannot directly predict individual student performance, cannot replace teachers' professional judgment, and still less can automatically trigger score adjustment. For real course evaluation, a more reasonable approach is to incorporate AI answering evidence, API review evidence with clearly bounded sources, student-group performance, problem-exposure variables, online-judge process data, item-setting structure over the years, course-level differences, and teacher interpretation together into a human-AI collaborative evaluation framework.

\hypertarget{references}{%
\section*{References}\label{references}}
\addcontentsline{toc}{section}{References}

{[}1{]} Miao G J, Liu Y, Xu N S, et al.~Research on the application of online judge systems in programming teaching{[}J{]}. Computer Education, 2016(09): 157-162. (in Chinese)

{[}2{]} Xue J, Chen R X, Zhang M, et al.~Design and implementation of an online judge teaching-assistance system for programming courses{[}J{]}. Computer Education, 2018(11): 104-108. (in Chinese)

{[}3{]} Cheang B, Kurnia A, Lim A, et al.~On automated grading of programming assignments in an academic institution{[}J{]}. Computers \& Education, 2003, 41(2): 121-131.

{[}4{]} Ihantola P, Ahoniemi T, Karavirta V, et al.~Review of recent systems for automatic assessment of programming assignments{[}C{]}//Proceedings of the 10th Koli Calling International Conference on Computing Education Research. New York: ACM, 2010: 86-93.

{[}5{]} Central Committee of the Communist Party of China, State Council. Overall plan for deepening educational evaluation reform in the new era{[}EB/OL{]}. Beijing: Central People's Government of the People's Republic of China, 2020{[}2026-06-25{]}. https://www.gov.cn/zhengce/2020-10/13/content\_5551032.htm. (in Chinese)

{[}6{]} Central Committee of the Communist Party of China, State Council. Outline of the plan for building a leading country in education (2024--2035){[}EB/OL{]}. Beijing: Central People's Government of the People's Republic of China, 2025{[}2026-06-25{]}. https://www.gov.cn/zhengce/202501/content\_6999913.htm. (in Chinese)

{[}7{]} Chen M, Tworek J, Jun H, et al.~Evaluating Large Language Models Trained on Code{[}J{]}. arXiv preprint arXiv:2107.03374, 2021.

{[}8{]} Austin J, Odena A, Nye M, et al.~Program Synthesis with Large Language Models{[}J{]}. arXiv preprint arXiv:2108.07732, 2021.

{[}9{]} Li Y, Choi D, Chung J, et al.~Competition-Level Code Generation with AlphaCode{[}J{]}. Science, 2022, 378(6624): 1092-1097.

{[}10{]} Messick S. Validity{[}M{]}//Linn R L, ed.~Educational Measurement. 3rd ed.~New York: Macmillan, 1989: 13-103.

{[}11{]} Kane M T. Validating the interpretations and uses of test scores{[}J{]}. Journal of Educational Measurement, 2013, 50(1): 1-73.

{[}12{]} Newton P E, Shaw S D. Validity in Educational and Psychological Assessment{[}M{]}. London: SAGE, 2014.

{[}13{]} American Educational Research Association, American Psychological Association, National Council on Measurement in Education. Standards for Educational and Psychological Testing{[}S{]}. Washington, DC: American Educational Research Association, 2014.

{[}14{]} Rasch G. Probabilistic Models for Some Intelligence and Attainment Tests{[}M{]}. Copenhagen: Danish Institute for Educational Research, 1960.

{[}15{]} Lord F M. Applications of Item Response Theory to Practical Testing Problems{[}M{]}. Hillsdale, NJ: Lawrence Erlbaum Associates, 1980.

{[}16{]} Siemens G. Learning analytics: The emergence of a discipline{[}J{]}. American Behavioral Scientist, 2013, 57(10): 1380-1400.

{[}17{]} Ferguson R. Learning analytics: Drivers, developments and challenges{[}J{]}. International Journal of Technology Enhanced Learning, 2012, 4(5/6): 304-317.

{[}18{]} Gašević D, Dawson S, Siemens G. Let's not forget: Learning analytics are about learning{[}J{]}. TechTrends, 2015, 59(1): 64-71.

{[}19{]} Pellegrino J W, Chudowsky N, Glaser R, eds.~Knowing What Students Know: The Science and Design of Educational Assessment{[}M{]}. Washington, DC: National Academies Press, 2001.

{[}20{]} Mislevy R J, Steinberg L S, Almond R G. On the structure of educational assessments{[}J{]}. Measurement: Interdisciplinary Research and Perspectives, 2003, 1(1): 3-62.

{[}21{]} UNESCO. Guidance for Generative AI in Education and Research{[}R{]}. Paris: UNESCO, 2023.

{[}22{]} Luckin R, Holmes W, Griffiths M, Forcier L B. Intelligence Unleashed: An Argument for AI in Education{[}R{]}. London: Pearson, 2016.

{[}23{]} Holmes W, Bialik M, Fadel C. Artificial Intelligence in Education: Promises and Implications for Teaching and Learning{[}M{]}. Boston: Center for Curriculum Redesign, 2019.

{[}24{]} Yang Z K, Cheng H, Wu L K. The basic framework of digital transformation in higher education and reflections on its practice{[}J{]}. Peking University Education Review, 2026(01): 57-71+193. (in Chinese)

{[}25{]} Jia J Y, Chen A X. Reasoning language models empowering education: practical value, risks and challenges, and development paths{[}J{]}. Peking University Education Review, 2026(01): 72-83+194. (in Chinese)

{[}26{]} Zhang Y, Hao Z X, Qin F. A systemic human-centered teaching philosophy leveraged by artificial intelligence{[}J{]}. Peking University Education Review, 2026(01): 84-96+195. (in Chinese)

{[}27{]} Cheng T J, Hong C. Cultural transformation and value examination of learning ethics in the digital-intelligence era{[}J{]}. Peking University Education Review, 2025(04): 101-116+187. (in Chinese)

{[}28{]} Zhu X D, Zhou J Y, Liu L S, et al.~Beyond academic achievement: constructing an assessment framework for Chinese children's all-round development and its support system{[}J{]}. Peking University Education Review, 2025(04): 117-140+188. (in Chinese)

{[}29{]} Becker B A, Denny P, Finnie-Ansley J, et al.~Programming Is Hard - Or at Least It Used to Be: Educational Opportunities and Challenges of AI Code Generation{[}C{]}//Proceedings of the 54th ACM Technical Symposium on Computer Science Education. New York: ACM, 2023: 500-506.

{[}30{]} Lau S, Guo P. From ``Ban It Till We Understand It'' to ``Resistance is Futile'': How University Programming Instructors Plan to Adapt as More Students Use AI Code Generation and Explanation Tools such as ChatGPT and GitHub Copilot{[}C{]}//Proceedings of the 2023 ACM Conference on International Computing Education Research - Volume 1. New York: ACM, 2023: 106-121. https://doi.org/10.1145/3568813.3600138.

{[}31{]} Finnie-Ansley J, Denny P, Luxton-Reilly A, et al.~The Robots Are Coming: Exploring the Implications of OpenAI Codex on Introductory Programming{[}C{]}//Proceedings of the 24th Australasian Computing Education Conference. New York: ACM, 2022: 10-19.

{[}32{]} Denny P, Prather J, Becker B A, et al.~Computing Education in the Era of Generative AI{[}J{]}. Communications of the ACM, 2024, 67(2): 56-67.

{[}33{]} Prather J, Denny P, Leinonen J, et al.~The Robots Are Here: Navigating the Generative AI Revolution in Computing Education{[}C{]}//Proceedings of the 2023 Working Group Reports on Innovation and Technology in Computer Science Education. New York: ACM, 2023: 108-159.

{[}34{]} Kasneci E, Sessler K, Küchemann S, et al.~ChatGPT for good? On opportunities and challenges of large language models for education{[}J{]}. Learning and Individual Differences, 2023, 103: 102274.

{[}35{]} Camilli G. Test fairness{[}M{]}//Brennan R L, ed.~Educational Measurement. 4th ed.~Westport, CT: Praeger, 2006: 221-256.

{[}36{]} Jiang Y, Chen J H, Wu Y C, et al.~Large language model-driven transformation of programming teaching{[}J{]}. Computer Education, 2025(08): 177-182. (in Chinese)

{[}37{]} Xu J J, Mao X G, Yin L Z, et al.~Restructuring the teaching of computer programming courses for large language models{[}J{]}. Computer Education, 2026(04): 156-161. (in Chinese)

\end{document}